\documentclass[11pt,a4paper]{article}
\pdfoutput=1 % if your are submitting a pdflatex (i.e. if you have images in pdf, png or jpg format)
\usepackage{jheppub}	% jheppub includes hyperref,color, natbib, amsmath, amssymb, epsfig, graphicx

\usepackage{tikz}
\usetikzlibrary{positioning,arrows}
\usetikzlibrary{decorations.pathmorphing}
\usetikzlibrary{decorations.markings}
\usetikzlibrary{snakes}
\usetikzlibrary{matrix}

\usepackage[utf8]{inputenc}
\usepackage[english]{babel}
\usepackage{makeidx}
\usepackage{tikz}
\tikzset{every picture/.style={line width=0.75pt}} %set default line width to 0.75pt
\usepackage{amsfonts}
\usepackage{enumerate}
\usepackage{mathrsfs}
\usepackage{tensor}
\usepackage[autostyle]{csquotes}
\usepackage{subfig}
\def\be{\begin{equation}}
	\def\ee{\end{equation}}
\def\bea{\begin{eqnarray}}
	\def\eea{\end{eqnarray}}
\newcommand{\nn}{\nonumber}

\def\apjl{\ref@jnl{ApJ}}

\usepackage{amsmath,amssymb}
\usepackage{latexsym}
\usepackage{graphicx}
\usepackage{slashed}

\usepackage[vcentermath,enableskew]{youngtab}
\usepackage{epsfig}

\def\be{\begin{equation}}
	\def\ee{\end{equation}}
\def\bea{\begin{eqnarray}}
	\def\eea{\end{eqnarray}}

\newcommand{\ba}{\begin{aligned}}
\newcommand{\ea}{\end{aligned}}

\title{\boldmath Partition functions of non-Lagrangian theories \\ from the holomorphic anomaly
}

\author[1]{Francesco Fucito,}
\author[2,3]{Alba Grassi,}
\author[1]{Jose Francisco Morales,}
\author[1]{and Raffaele Savelli}
\affiliation[1]{
	I.N.F.N -- sezione di Roma Tor Vergata and Dipartimento di Fisica\\
	Via della Ricerca Scientifica, I-00133 Roma, Italy
}

\affiliation[2]{
Section de Math\'ematique, Universit\'e de G\'eneve,
1211 Geneva 4, Switzerland}
\affiliation[3]{
CERN, Theory Division,
1211 Geneva 23, Switzerland
}

\makeatletter
\renewcommand{\@email}[1]{#1}
\makeatother

\emailAdd{fucito@roma2.infn.it}
\emailAdd{alba.grassi@cern.ch}
\emailAdd{morales@roma2.infn.it}
\emailAdd{savelli@roma2.infn.it}

\abstract{The computation of the partition function in certain quantum field theories, such as those of the Argyres-Douglas or Minahan-Nemeschansky type, is problematic due to the lack of a Lagrangian description. In this paper, we use the holomorphic anomaly equation to derive the gravitational corrections to the prepotential of such theories at rank one by deforming them from the conformal point. In the conformal limit, we find a general formula for the partition function as a sum of hypergeometric functions. We show explicit results for the round sphere and the Nekrasov-Shatashvili phases of the $\Omega$ background. The first case is relevant for the derivation of extremal correlators in flat space, whereas the second one has interesting applications for the study of anharmonic oscillators.
}

\preprint{CERN-TH-2023-095}

\begin{document}

\maketitle

\section{Introduction}

In \cite{Argyres:1995jj,Argyres:1995xn} it was shown that in a suitable limit, the Argyres-Douglas (AD) limit, the moduli space of a massive $\mathcal{N}=2$ supersymmetric gauge theory of the Yang-Mills type leads to isolated superconformal field theories (SCFT). A first attempt to classify such theories appeared in \cite{Eguchi:1996ds,Eguchi:1996vu}
while more recent results were obtained in \cite{Argyres:2015ffa,Argyres:2015gha,Argyres:2016xua,Argyres:2016xmc,Chacaltana:2014nya,Martone:2021ixp,Xie:2012hs,Cecotti:2012jx,Cecotti:2013lda,Wang:2015mra}. In this paper we will be concerned with the rank-one version of the AD SCFT's as well as of those of the Minahan-Nemeschansky (MN) type \cite{Minahan:1996cj}. Being such SCFT's isolated and strongly coupled, their analytic treatment is troublesome given that a Lagrangian description is not available. To circumvent these difficulties,  at least five different strategies have appeared in the literature: The conformal bootstrap, the AGT duality, the matrix-model methodology,  the large-charge expansion, and the geometric approach based on the $\Omega$-background technology. The numerical conformal bootstrap is a method that exploits the constraints coming from the symmetries of the theory to give numerical estimates of the parameters of interest \cite{Rattazzi:2008pe,Beem:2014zpa,Cornagliotto:2017snu,Gimenez-Grau:2020jrx,Simmons-Duffin:2016gjk}.
The AGT duality, in its original formulation, relates the partition function of a four-dimensional SQCD with four massive flavors with a four-point correlator of a two-dimensional conformal field theory \cite{Alday:2009aq,Wyllard:2009hg}. Making some (or all) of these points collide leads to rank-one SCFT's \cite{Gaiotto:2012sf,Bonelli:2011aa,Bonelli:2016qwg,Kimura:2020krd,Nagoya:2015cja,Nagoya:2018pgp}. Similar ideas have also been  used to provide matrix-model representations for the partition function in a class of  AD  theories \cite{Nishinaka:2012kn,Rim:2016ldj, Itoyama:2018gnh,Grassi:2018spf,Nishinaka:2019nuy,Itoyama:2019rgp,Oota:2021qky,Itoyama:2021nmj}.  Both the AGT and the matrix-model technology have been useful to study AD theories with large deformation parameters (see also \cite{Fucitonew}). Another original perspective has been explored  in the context of the large-charge expansion \cite{Hellerman:2015nra,Bourget:2018obm,Monin:2016jmo,Alvarez-Gaume:2016vff,Jafferis:2017zna,Hellerman:2017sur,Hellerman:2018xpi,Grassi:2019txd,Beccaria:2020azj}, where it was suggested that one can have an approximate description of such strongly coupled SCFT's in terms of an universal effective field theory. Finally, using the genus expansion of the $\Omega$ background, as well as ideas coming from   localization \cite{Losev:1997tp,Moore:1997dj,Lossev:1997bz,Nekrasov:2002qd,Flume:2002az,Bruzzo:2002xf,Nekrasov:2003rj,Pestun:2007rz,mpp,Fucito:2016jng,Gerchkovitz:2016gxx,Rodriguez-Gomez:2016cem,Rodriguez-Gomez:2016ijh,Billo:2017glv,Billo:2019job}, it has been possible to study chiral/anti-chiral correlators of non-Lagrangian theories \cite{Grassi:2019txd,Bissi:2021rei}.
Such analytic results, even if based on the first two leading terms in the expansion of the prepotential for small curvatures,  show surprisingly good agreement with the numerical bootstrap method \cite{Cornagliotto:2017snu} as well as with the large-charge expansion \cite{Hellerman:2017sur,Hellerman:2018xpi}.  To improve the  analytic estimate of \cite{Grassi:2019txd,Bissi:2021rei}  and get an exact result, one should incorporate higher curvature terms in the prepotential. This is the main motivation of this paper.

To accomplish this task we use the recursion equations following from the refined holomorphic anomaly. The latter was originally investigated in topological field theories \cite{Bershadsky:1993ta}, and then  revisited in \cite{Huang:2006si,Huang:2009md,Huang:2011qx,Huang:2013eja,Billo:2013fi,Billo:2013jba,Ashok:2015cba,Billo:2015pjb,Billo:2015jyt, Krefl:2010fm,Codesido:2017dns,Fischbach:2018yiu} after the introduction of the $\Omega$  background \cite{Losev:1997tp,Moore:1997dj,Lossev:1997bz,Nekrasov:2002qd,Flume:2002az,Bruzzo:2002xf}. We want to emphasize that, when approaching the AD point, it is essential to employ the holomorphic anomaly equation. Indeed this technique, while providing expressions which are perturbative in the  $\Omega$-background parameters, is exact in all the other parameters of the theory. This is an important difference with respect to localization techniques \`a la Nekrasov, which instead cannot be used in the context of strongly coupled field theories.

All rank-one SCFT's of the AD and MN type are characterized by the dimension of their Coulomb-branch parameter and they can be treated in a uniform way.
First, we specialize the holomorphic anomaly equation
to a specific one-parameter family of deformations of these SCFTs.  This allows us to
 compute the free energy exactly in the deformation parameters and order by order in the $\Omega$ background parameters $\epsilon_{1,2}$.  When we turn off the deformation and go to the conformal point, we discover significant simplifications.
More precisely, we find that their partition function can be expressed as an infinite sum of confluent U-hypergeometric functions
\be
 {\cal Z}(a,\epsilon_1,\epsilon_2) =  {\rm e}^{{\cal F}_0(a)  \over \epsilon_1 \epsilon_2 } \, E_2^{\gamma /2}\sum_{n=0}^\infty\left(\frac{E_{2\delta}}{E_2^{\delta}}\right)^n c_n \,{\rm U}\left(-\frac{\gamma}{2}+n\delta,\frac{1}{2},-\frac{6a^2}{E_2\epsilon_1\epsilon_2}\right)\,\quad {\epsilon_1 \epsilon_2\neq 0}\,,
 \label{zuin}    \ee
where $a$ is the local coordinate on the Coulomb branch, $\delta=2\,,3$ depending on the SCFT, $E_I$ are the Eisenstein functions evaluated at the fixed value of the modular parameter $\tau_*={\rm i}\,, e^{\pi {\rm i} \over 3}$,  and $\gamma$ a constant determined by the conformal dimension of the Coulomb-branch operator (see Sec.~\ref{Sec:PartFunct} for more details). Finally, the coefficients $c_n$ are pure rational numbers depending only on the phase of  the $\Omega$ background. They are determined by the gap conditions \cite{Huang:2011qx,Krefl:2010fm}, that ensure consistency of the expansion near singular monopole points. In order to compute them, deforming away from the conformal point is essential. Nevertheless, we check that their value is independent of the particular deformation we choose.
We will also show that in the so-called Nekrasov-Shatashvili limit (NS) \cite{Nekrasov:2009rc}, i.e.~$\epsilon_1\to 0$, the summation in  \eqref{zuin} undergoes a non-trivial re-organization in terms of a different set of functions. This limit is relevant for the study of
quantum-mechanical anharmonic oscillators.

This paper is organized as follows.
In Section \ref{Sec:Review} we review the holomorphic anomaly equation and explain how to solve it recursively. In Section \ref{Sec:General} we specialize this algorithm to the isolated rank-one conformal field theories and show that important simplifications occur leading to \eqref{zuin}.
In Section \ref{Sec:Sphere} we focus on  the example of the sphere ($\epsilon_1=\epsilon_2$) which is relevant for the computation of the extremal correlators of these SCFTs.
In Section \ref{Sec:NS} we discuss the NS limit. We conclude in Section \ref{Sec:Outlook} with a few hints for further investigations. Several technical details as well as conventions are relegated to five appendices.

\section{The $\Omega$-background prepotential}\label{Sec:Review}

\subsection{Holomorphic anomaly equation}

We consider rank-one ${\cal N}=2$ supersymmetric (in general non-Lagrangian) theories living on an $\Omega$-background specified by the parameters $\epsilon_1,\epsilon_2$ and by a Seiberg-Witten (SW)  geometry.   We denote by $(a,a_D)$ the SW periods, by $u$ the Coulomb-branch parameter and omit the dependence on all remaining parameters: couplings and masses.
The partition function  on the $\Omega$-background can be written as
\be
{\cal Z}(a,\epsilon_1,\epsilon_2) = e^{ {\cal F}(a,\epsilon_1,\epsilon_2) \over \epsilon_1 \epsilon_2}
\ee
with ${\cal F}$ the prepotential. The theory prepotential is regular in the limit $\epsilon_1,\epsilon_2\to 0$ so it can be expanded as
\be\label{Fge1e2}
{\cal F}(a,\epsilon_1,\epsilon_2)  = \sum_{g=0}   (\epsilon_1 \epsilon_2)^g {\cal F}_g(a,\beta) =\sum_{h,s\geq 0} (\epsilon_1+\epsilon_2)^{2h} (\epsilon_1 \epsilon_2)^{s} {\cal F}_{s,h}(a)
\ee
 with
\bea
{\cal F}_g(a,\beta) &=&\sum_{h=0}^g (\beta+\beta^{-1})^{2h}   {\cal F}_{g-h,h}(a) \qquad , \qquad \beta = \sqrt{\epsilon_1 \over \epsilon_2 }~.
\eea
The ${\cal F}_0(a)$ term represents the theory prepotential in flat space which can be determined out of the SW geometry. Higher derivative terms are given by the reduced partition function
\be
\widehat{\cal Z}(a,\epsilon_1,\epsilon_2)=e^{-{{\cal F}_0(a)\over \epsilon_1 \epsilon_2} }{\cal Z}(a,\epsilon_1,\epsilon_2)
\ee
that unlike ${\cal Z}$ has a regular limit when the $\Omega$-background is turned off. This function will be the main object of our study.
We introduce the IR coupling
\bea\label{coupling}
q (a)& =& e^{\pi {\rm i} \tau (a)}=  e^{\pi {\rm i} {\partial a_D\over \partial a} } =e^{-{\partial^2 {\cal F}_0 (a)\over 2 \partial a^2} }~.
\eea
 The partition function $ \widehat{\cal Z}$ can be alternatively viewed  as a function of $q$ or as a function of $a$.  In particular, one can express $\widehat{\cal Z}(q)$  in terms of the Eisenstein's series $E_2(q),E_4(q),E_6(q)$ that form a basis of quasi-modular functions, see App.~\ref{EisensteinApp} for all the relevant definitions.\footnote{As we will see later, in specific cases it is convenient to replace $E_4(q),E_6(q)$ with different modular functions. } All $\mathcal{F}_g(q)$'s have weight zero and $a(q)$ has weight one. S-duality covariance constrains the dependence of the partition function on $E_2$. Indeed the full dependence on  this form is determined by the anomaly equation \cite{Witten:1993ed,Huang:2011qx,Billo:2013jba,Krefl:2010fm}\footnote{Throughout this paper we consider the holomorphic version of the anomaly equation, obtained by replacing $\widehat{E}_2(\tau, \overline{\tau})=E_2(\tau)-{3\over \pi {\rm Im}(\tau)}\to E_2(\tau)$.}
  \be\label{heat}
\partial_{E_2} \widehat{\cal Z} (q)  = \frac{\epsilon_1 \epsilon_2}{24}  \,\partial^2_a  \widehat{\cal Z}(q)  \,~.
\ee
In (\ref{heat}) the derivatives on the l.h.s.~is carried out keeping $E_4$, $E_6$ constant.
In the r.h.s. of (\ref{heat}) the partition functions is meant as a function of $q$ and $a$.
Writing
\be
 \widehat{\cal Z}(q,\epsilon_1,\epsilon_2) = \sum_{g=0}^\infty (\epsilon_1 \epsilon_2)^g \widehat{\cal Z}_g (q,\beta)
\ee
one finds the recursive equation
 \be
\partial_{E_2} \widehat{\cal Z}_g   = \frac{1}{24}  \,\partial^2_a  \widehat{\cal Z}_{g-1}  \,.\label{heat2}
\ee
  For the ``reduced'' prepotential $\widehat{\cal F}(q,\beta)={\cal F}(q,\beta) -{\cal F}_0(q)$ one finds
\be   \label{hafh}
\partial_{E_2} \widehat{\cal F}= \frac{1}{24}  \left[ \epsilon_1 \epsilon_2\partial^2_a  \widehat{\cal F} +\left(  \partial_a  \widehat{\cal F} \right)^2 \right]  \,, \ee
  or equivalently, using \eqref{Fge1e2},
    \be\label{HAE}
 \partial_{E_2} {\cal F}_{g} = \frac{1}{24} \left[    \partial_a^2 {\cal F}_{g-1} +\sum_{g'=1}^{g-1} \partial_a {\cal F}_{g'} \partial_a {\cal F}_{g-g'}  \right] ~.
\ee
Equations (\ref{HAE}) allows to compute ${\cal F}_g$ recursively starting from ${\cal F}_1(q,\beta)$ up to $E_2$-independent terms.
   On the other hand ${\cal F}_1(q,\beta)$
 is determined in terms of ${\partial a\over \partial u}(q)$ and the discriminant $\Delta(q)$  characterising the dynamics in flat space via the formula
\be
{\cal F}_1 (q,\beta) =-{1 \over 2}  \log {\partial a\over \partial u}(q) + \frac{\beta^2+\beta^{-2} }{24}\log \Delta (q)~.
\ee

 \subsection{Seiberg-Witten elliptic curve}

 The functions ${\partial a\over \partial u}(q)$ and $\Delta(q)$ entering ${\cal F}_1$ are described by the SW elliptic geometry. For a recent discussion see also \cite{Aspman:2021vhs}.
 We write the Seiberg-Witten curve in the Weierstrass form\footnote{In App.~\ref{App:SCQDtoAD}, we collect some results which are useful to bring to this standard Weierstrass form the different expressions used in the literature for the elliptic geometry of rank-one theories.}
\be
y^2=4 z^3-g_2(u)  z-g_3(u)\,,
\ee
with discriminant
\be
\Delta (u)=16[g_2^3(u)-27g_3^2(u)]\,.
\ee
The SW periods are given by
  \bea
 \omega_1 ={\partial a  \over \partial u} &=&  {1\over \pi}  \oint_{\alpha}  {dz\over y(z) }
 \qquad  , \qquad
w_2 ={\partial a_D  \over \partial u} =  {1\over \pi}   \oint_{\beta}  {dz\over y(z) }
 \label{periods}
\eea
and the complex coupling parameter $q$  introduced in (\ref{coupling}) can also be given in terms of
\be
q=e^{ \pi {\rm i} w_2/ \omega_1}~.
\ee
The functional dependence $u(q)$ and $ \omega_1(q)$  is determined by solving
  the elliptic geometry formulae
\be
g_2(u) ={4 E_4(q)\over 3  \omega_1(q)^4}  \quad, \qquad g_3 (u) ={8E_6(q) \over 27  \omega_1(q)^6}  \label{unu}
\ee
for $u(q)$ and $ \omega_1(q)$ in terms of $E_4(q)$ and $E_6(q)$.
Once this is done, all functions of $u$ can be viewed as functions of $q$. For example, the discriminant is given by
\be\label{Discriminantg2g3}
\Delta (q) = 16 (g_2^3-27 g_3^2) ={1024(E_4(q)^3-E_6(q)^2)\over 27  \omega_1(q)^{12}} %= { 16^4 \, \eta^{24} \over  \omega_1^{12} }
\ee
and the first gravitational correction becomes
\be\label{F1general}
 {\cal F}_1 (q)=-{1 \over 2}  \log  \omega_1(q) + \frac{\beta^2+\beta^{-2} }{24}\log \Delta (q)~.
 \ee
  To compute higher derivative terms, we need derivatives with respect to $a$, that can be translated into derivatives with
 respect to $q$ using the chain rule
\be
  \partial_a   {\cal F}_g (q,\beta) =   \xi   D_\tau  {\cal F}_g(q,\beta)    \label{dFsummary0}
 \ee
with
 \be
 D_\tau= {\partial_\tau\over \pi {\rm i} } =q \partial_q
 \ee
 and
 \be
 \xi =  q^{-1}{dq \over da } = {1\over  \omega_1 D_\tau u}   =  { \nu'(u)  \, E_4 \, E_6 \over 2  \omega_1  (E_6^2-E_4^3) } \label{xigeneral}
 \ee
 where $\xi$ is a modular form of weight $-3$ and
 \be
  \nu(u)=\log{ 27 g_3(u)^2\over g_2(u)^3}=\log{ E_6(q)^2\over E_4(q)^3} \,.\label{nuu0}
 \ee
 Here we have used (\ref{unu}) and computed the derivatives with respect to $\tau$ using
 \be
 D_\tau E_2 =  \frac{1}{6} (E_2^2-E_4) \quad  , \quad  D_\tau E_4 =  \frac{2}{3} (E_2 \, E_4-E_6)  \quad , \quad
  D_\tau E_6=  E_2 \, E_6-E_4^2\,.
 \ee
       Plugging (\ref{dFsummary0}) into (\ref{HAE}) one can compute ${\cal F}_g$ order by order in $g$ up to $E_2$ invariant terms. The general form of $\mathcal{F}_g$ is
    \be \label{Fggen}
    \mathcal{F}_g(q,\beta)=\xi^{2g-2}\left(\sum_{\ell=1}^{3g-3} c_{g,\ell}(\beta,E_4,E_6) \,E_2^\ell  +h_g (\beta,E_4,E_6)\right) ,\quad g\geq 2\,,
\ee
    where $c_{g,\ell}(\beta,E_4,E_6)$ is a modular form of weight $6g-6-2\ell$ and $h_g(\beta,E_4,E_6)$ is a modular function of weight $6g-6$,  known as holomorphic ambiguity, which cannot be determined using (\ref{HAE}).

    The holomorphic ambiguities are fixed by imposing the so-called ``gap conditions'' \cite{Huang:2009md,Hori:2003ic,Krefl:2010fm,Huang:2011qx}, that determine the behavior of the prepotential near the points where the elliptic curve degenerates.  Rank-one SCFT's can be deformed in such a way that the discriminant of their SW curve takes the particularly simple form
   \be\label{Deltasimp}
   \Delta(u)  \sim \prod_{i=1}^{n} (u-u_0 e^{2\pi {\rm i} \over n} )= u^n-u_0^n
   \ee
   leading to $n$ equivalent singularities in the $u$-plane.

According to (\ref{Discriminantg2g3}), the zeroes of the discriminant $u\sim u_0$ correspond to the point where $q=0$. This limit can be studied, expanding
   \be
   a(q)  =\int^q { dq' \over q' \xi(q') }
   \ee
    for small $q$, and inverting the series to get $q(a)$ for small $a$. Plugging this into the ${\cal F}_g$, the holomorphic ambiguities $h_g$ are determined by
   requiring the gap conditions  \cite{Krefl:2010fm, Fischbach:2018yiu}
  \bea\label{GeneralBernoulli}
  {\cal F}_g ( a)   & \underset{ q\to 0} {\approx}  &  (2g-3)!
  \sum_{k=0}^{g} \widehat{B}_{2k}\widehat{B}_{2g-2k}{ \beta^{2g-2k}\over a^{2g-2} } + O( a^0)
  \eea
 where \be\widehat B_ {m}=\frac{\left(\frac{1}{2^{m-1}}-1\right) B_m}{m!} \ee and $B_k$ are the Bernoulli numbers. We will work out  two different choices of $\Omega$ background, i.e.~$\beta=1$  and $\beta=0$. In these cases \eqref{GeneralBernoulli} becomes:
 \bea \label{sph}
 \beta=1: & &{\cal F}_g ( a)    \underset{ q\to 0} {\approx}    -{   \frac{B_{2g}} {2 g(2g-2)\,a^{2g-2}} }  + O( a^0) \,, \label{Bernoulli1}\\
 \label{gapns}
 \beta=0: & &  {\cal F}_g ( a) \underset{ q\to 0} {\approx}  - \frac{\left(1-2^{1-2 g}\right) (2 g-3)! B_{2 g}}{(2 g)! \,a^{2g-2}}  + O( a^0)\,.\label{Bernoulli0}
 \eea
 It is important to stress that in more complicated setups,\footnote{For instance where \eqref{Deltasimp} does not hold.} equations (\ref{unu}) or (\ref{nuu0}) are hard to solve or admit  several inequivalent solutions, often related to each other by modular transformations. In such cases (see App.~\ref{cDefH1} for details), the ${\cal F}_g$'s transform
 non-trivially under modular transformations and a different set of gap conditions on their modular transformed ${\cal F}_g^D$ is required at $q_D \to 0$. The basis of modular functions to use is also adapted according to such situations.

\section{Isolated rank-$1$ conformal field theories}\label{Sec:General}

\subsection{The partition function}\label{Sec:PartFunct}

\begin{table}
\centering
$
 \begin{array}{|c|cccccc|}
\hline
 {\rm SW} & \mathcal{H}_0 &  \mathcal{H}_1 &  \mathcal{H}_2 &  {\rm E}_6 &  {\rm E}_7 &  {\rm E}_8 \\
 \hline
 N_{7} & 2 & 3 & 4&   8 & 9 & 10\\
 d  & \frac{6}{5} & \frac{4}{3} & \frac{3}{2} &   3 & 4 & 6 \\
 % \alpha &2& 3 & 4 &  8 &9& 10\\
  g_2 & 0 & u & 0  &  0 & u^3 & 0\\
  g_3 & u & 0& u^2   & u^4 & 0 & u^5\\
  \tau & e^{\pi {\rm i} \over 3} & {\rm i} &  e^{\pi {\rm i} \over 3}  &  e^{\pi {\rm i} \over 3}& {\rm i} & e^{\pi {\rm i} \over 3}\\
\hline
 \end{array}
$
\caption{SW data for isolated rank-1 ${\cal N}=2$ SCFTs}\label{tSW}
\end{table}

Rank-one conformal field theories  can be realized in F-theory as a single D3-brane, probing a singularity built out of a certain number $N_7$ of coinciding mutually non-perturbative 7-branes \cite{Banks:1996nj}. The low-energy dynamics on the D3-brane is described by a SW elliptic curve specified by a single Coulomb-branch parameter $u$, a $u$-independent modular parameter $\tau$  and a discriminant
\be
\Delta(u) \sim g_2^3-27 g_3^2 \sim u^{N_7 }~. \label{discun}
\ee
Prototypical examples are the AD theories $\mathcal{H}_0$, $\mathcal{H}_1$, $\mathcal{H}_2$, and the Minahan-Nemeschansky theories E$_6$, E$_7$, E$_8$. These are all isolated, non-Lagrangian field theories and are the focus of the present work.
They can be split into two classes depending on the value of the modular parameter
\bea\label{Classification}
{\bf A}: && \quad   \tau=e^{\pi {\rm i} \over 3} \quad , \quad y^2=4x^3-u^{b_3} \quad, \quad b_3=1,2,4,5 \quad, \quad  \mathcal{H}_0,\mathcal{H}_2,  {\rm E}_6, {\rm E}_8\nn\\
{\bf B}: && \quad  \tau= {\rm i} \quad , \quad ~~~y^2=4x^3-u^{b_2} x ~~, \quad b_2=1,3 \quad, \qquad  ~~~~~  \mathcal{H}_1, {\rm E}_7
\eea
   The conformal dimension $d$ of the Coulomb-branch parameter is given by
 \be
d= {12 \over 12 - N_7} \label{du}
 \ee
  that follows from the requirement that the SW period $ \partial a\over \partial u $ be of dimension $1-d$ and therefore the conformal dimension of the holomorphic differential be  $\left[ dx/y\right]=1-d$. From (\ref{discun}) and (\ref{du}) it follows that $N_7$ is an integer,
  multiple of 2 or 3, and smaller than 12.
    In  Table~\ref{tSW} we collect the SCFT data for all possible choices of $N_7$.\footnote{The case of $N_7=6$ is special because both $g_2$ and $g_3$ are generically non-vanishing, with the ratio $g_2^3/g_3^2$ an arbitrary complex number. The associated SCFT is therefore not isolated and it corresponds to the $SU(2)$ gauge theory with four massless fundamental hypermultiplets.}
In the theories of type {\bf{A}} the modular form $E_4$ vanishes, whereas $E_2, E_6$ are constants.
Similarly in the theories of type {\bf{B}} the modular form $E_6$ vanishes, whereas $E_2, E_4$ are constants. Therefore, in all these cases, the free energy is a function of $\beta$ and of the following dimensionless quantities
\bea\label{xtoa}
 x=\frac{E_2\epsilon_1\epsilon_2}{6a^2} \qquad \qquad \kappa=\frac{E_{2\delta}}{E_2^{\delta}}\,,
\eea
where
\be
\delta=\left\{\begin{array}{lc}3 & \qquad{\bf A}\\ 2 &\qquad {\bf B} \end{array}\right. ~.
   \ee
 For these SCFTs one finds
  \be
  u \sim a^d \qquad , \qquad \Delta(u)=a^{12 (d-1)} \,,
  \ee
  where $d$ is the conformal dimension of the Coulomb-branch operator (see Table~\ref{tSW}).
    The first correction to the SW prepotential takes the general form\footnote{Throughout the paper we will omit any additive constant to $\mathcal{F}_1$.}
    \be
{\cal F}_1 (a,\beta) = \gamma \log  \left({a\over {\sqrt{\epsilon_1\epsilon_2}}}\right)
\ee
 with
 \be
 \gamma = \frac{d-1}{ 2} (1+\beta^2+\beta^{-2})\,.
 \ee
 By dimensional analysis, the higher corrections take the form
 \be\label{Fad}
 \mathcal{F}_g(a,\beta)= \frac{{\mathfrak{f}_g}(\beta)}{a^{2g-2}}\,,
 \ee
 where $\mathfrak{f}_g(\beta)$ are numbers. The latter can be computed recursively using the holomorphic anomaly equation  with boundary conditions fixed by an $E_2$-independent function. We can make the following Ansatz
\be
 \widehat{\cal Z}(a,\beta) = E_2^{\tfrac{\gamma}2}\sum_{n=0}^\infty\kappa^n c_n f_n(x,\beta)\,,
\label{zansatz}\ee
%\be \label{zansatz2}  \widehat{\cal Z}(a,\beta) = \left({a\over  {\epsilon_1\epsilon_2} }\right)^\gamma\sum_{n=0}^\infty\kappa^n c_n(\beta) f_n(x,\beta)\,,
%\ee
where $c_n$ are numerical coefficients encoding the holomorphic ambiguities and depend on the phase of the $\Omega$ background.
Plugging (\ref{zansatz}) into \eqref{heat}
 leads to the confluent hypergeometric equation\footnote{We remark that in a SCFT $\tau$ is independent of $a$, and thus $E_2$ and $a$ are independent variables.}
 \be
2x^3 f_n''(x)+x\left(3x-2\right) f_n'(x)+\left(2n\delta-\gamma\right)  \,f_n(x)=0\,.
\ee
where the boundary conditions are chosen such that \eqref{zansatz} has a power-like behavior for $x\to0$. The final solution is
 \be  \label{zu}{
 \widehat{\cal Z}(a,\beta) = E_2^{\gamma /2}\sum_{n=0}^\infty\kappa^n c_n \,{\rm U}\left(-\frac{\gamma}{2}+n\delta,\frac{1}{2},-\frac{1}{x}\right)\,, }
 \ee
  with U$(a,b,z)$ the confluent $U$ hypergeometric function\footnote{Our conventions are the same as in {\tt Mathemetica}.}  and $c_0$ is an overall normalization  which can be set to  $c_0=1$ without loss of generality.
    The coefficients $\{c_n\}_{n\geq 1}$ are $\beta$-dependent coefficients encoded in the $E_2$-independent part of $\widehat{\cal Z}$. In the next section we will derive the first few coefficients $c_n$ for the theories in Table~\ref{tSW}, and show that they are rational numbers. The strategy will be to first turn on suitable mass or coupling deformations for such theories, in order to isolate a monopole point where the gap condition can be imposed. The coefficients $c_n$ will then be derived by turning off the deformation.\footnote{In \cite{Moore:2017cmm} it was also observed that, to determine the  partition function of topologically twisted $\mathcal{H}_0$, one has to first  perturb the theory away from the conformal point.} We  check explicitly that the final result is independent of the deformation. We also note that \eqref{zu} as it is written holds for $\epsilon_i\neq 0$, i.e.~all phases of the $\Omega$ background except the NS phase. Indeed if we consider the NS limit there is a non-trivial re-organization of  \eqref{zu} which we discuss in Sec.~\ref{Sec:NS}.

\subsection{Deformations}

Conformal invariance can be broken by turning on masses or couplings. Here we consider the simplest deformation splitting democratically the discriminant into its $N_7$ roots
\be
\Delta(u) \sim u^{N_7} -m^{d N_7}
\ee
We will refer to $m$ generically as a mass deformation, although for the case of ${\cal H}_0$, where masses are not available,
the dimension-one parameter $m$ is related to the IR-relevant coupling $c$ via $m=c^{5\over 4}$. The deformed SW curves look like

\bea\label{Classification2}
{\bf A}: &&   \quad \quad y^2=4x^3-m^{ 4 b_3  \over 6 -b_3   } x -u^{b_3} \quad, \quad b_3=1,2,4,5  \nn\\
{\bf B}: && \quad   \quad y^2=4x^3-u^{b_2} x-m^{ 6 b_2 \over 4 -b_2}  ~~, \quad b_2=1,3
\eea
In all these examples, we will derive $q$-exact formulae for the first few $\mathcal{F}_g$'s. An important ingredient in our procedure will be to parametrize the holomorphic ambiguities for {\bf A} and {\bf B} theories respectively in the following form ($g\ge2$)
   \bea\label{iminmax}
   h^A_g(\beta,q)& =&\frac{E_4^{3g-3}}{E_6^{g-1}} \sum_{i=0}^{\left[{5g-5}\over{3}\right]} \left(\frac{E_6^2}{E_4^3}\right)^i  h_{g,i}(\beta) ~, \nn\\
   h^B_g(\beta,q) &=&  E_6^{g-1} \sum_{i=0}^{\left[ {3g-3\over 2} \right] } \left(\frac{E_4^3}{E_6^2}\right)^i  h_{g,i}(\beta)  ~,
   \eea
where $h_{g,i}(\beta)$ are $q$-independent  coefficients to be determined. The above expressions are dictated by the requirement that $h_g$ has modular weight $6g-6$, allowing only integer powers of $E_4$ and $E_6$, such that $h_g$ does not grow faster than its corresponding non-ambiguous part when $E_4\to0$ and $E_6\to0$.

 In App.~\ref{cDefH1}, we will consider an alternative deformation
of the AD theory  ${\cal H}_1$  described by the SW curve
   \be
   y^2=4 x^3- u x - cu+4c^3
   \ee
   where $c$ is the IR-relevant coupling. In particular we will show that the results for the $\mathcal{F}_g$'s in the conformal limit are the same, independently of the deformation used to compute them.
An analogous match for the theory ${\cal H}_2$ is obtained in App.~\ref{SQCDNf3}, where we consider the $N_f=3$-SQCD description of this AD theory.

 \section{Examples: $\beta=1$}\label{Sec:Sphere}

 In this section, we consider the $\Omega$ background given by $\epsilon_1=\epsilon_2=\epsilon$, i.e.~$\beta=1$. This choice enters for example the computation of the round-sphere partition function \cite{Pestun:2007rz,Okuda:2010ke} and of extremal correlators \cite{Gerchkovitz:2016gxx,Rodriguez-Gomez:2016cem,Rodriguez-Gomez:2016ijh,Billo:2017glv,Grassi:2019txd,Beccaria:2020azj,Bissi:2021rei}. Despite such a large interest, the holomorphic anomaly techniques have not been explored so far for this particular phase of the $\Omega$ background.\footnote{We also note that the holomorphic anomaly equation for the sphere and the standard topological string phase ($\epsilon_1=-\epsilon_2$) is actually the same. What changes are the initial data, i.e.~$\mathcal{F}_1$, and the gap conditions.} In the following we will compute  \eqref{Fggen} stopping at the first order in $g$ in which the holomorphic ambiguity contributes in the conformal limit. This is dictated by a reason of simplicity given that the formulae become very large. In App.~\ref{AppE} we will give results up to $g=18, 7, 15$ for $\mathcal{H}_0, \mathcal{H}_1, \mathcal{H}_2$ respectively.

 \subsection{$\mathcal{H}_0$ theory}

 The SW curve for the deformed $\mathcal{H}_0$ theory is
 \be
 y^2=4x^3-cx-u\,.
 \ee
Plugging $g_2=c$, $g_3=u$ into (\ref{unu}) and \eqref{xigeneral}  gives
 \bea\label{formulaeH0}
 u &=&  {c^{3\over 2}  E_6(q)\over 3\sqrt{3} E_4(q)^{3\over 2} } \qquad , \qquad
  \omega_1 =\left( {4 E_4\over 3 c}\right)^{1\over 4}
  \qquad , \qquad  \xi ={  3^{7\over 4} \, E_4(q)^{9\over 4} \over  2^{1\over 2}  c^{5\over 4} (E_6^2-E_4^3) } \nn\\
 {\cal F}_1 &=&  \frac{1}{12} \log
   \left( c^{9\over 2} \frac{E_4(q)^3 - E_6(q)^2}{E_4(q)^{9\over 2} }\right)   \quad , \quad \Delta=16(c^3-27 u^2) ~~.
 \eea
 In this case the holomorphic ambiguity takes the form of the first expression in (\ref{iminmax}).
% \be
%i_{\rm min} =\left[{1-g\over 2} \right] \qquad, \qquad i_{\rm max} =\left[{3(g-1)\over 2} \right]
% \ee
 Solving recursively the holomorphic anomaly equation (\ref{HAE}),  one finds the first few terms:
 {\small
  \bea\label{FgH0cu}
&&  {\cal F}_2 =  {  \xi^2\over 24 \, 12^2}  \left[  \frac{5}{3}E_2^3 +\frac{ 3  E_6 }{ E_4}E_2^2 -\frac{\left( 34 E_4^3 +21
   E_6^2\right) }{ E_4^2} E_2   + h_2(q) \right] \nn \\
 &&  {\cal F}_3 = { \xi^4\over 24 \, 12^4}  \left[  \frac{5 }{6}E_2^6+\frac{10 E_6 }{E_4}E_2^5+\frac{\left(16
   E_4^3+67 E_6^2\right) }{2 E_4^2}E_2^4 -\frac{\left(1465
   E_6 E_4^3+147 E_6^3\right) }{9
   E_4^3}E_2^3 \right.  \nn\\
   && \quad \left.  -\frac{\left(11897 E_4^6+59376 E_6^2 E_4^3+6300
   E_6^4\right) }{30 E_4^4}E_2^2 +\frac{\left(104257 E_6
   E_4^3+95565 E_6^3\right) }{15 E_4^2}E_2+h_3(q) \right]  \nn\\
   &&  {\cal F}_4 = { \xi^6\over 24 \, 12^6}  \left[    \frac{1105 }{1296}E_2^9+\frac{865 E_6 }{48 E_4}E_2^8+\left(\frac{3589 E_6^2}{24 E_4^2}+\frac{2039 E_4}{72}\right)
   E_2^7+\left(\frac{41491 E_6^3}{72 E_4^3}+\frac{69869 E_6}{216}\right) E_2^6 \right.\nn\\
&&\left.
   +\left(\frac{175987 E_6^4}{240
   E_4^4}-\frac{43813 E_6^2}{60 E_4}-\frac{149791 E_4^2}{720}\right) E_2^5-\left(\frac{76559 E_6^5}{48
   E_4^5}+\frac{399439 E_6^3}{15 E_4^2}+\frac{10250789 E_4 E_6}{720}\right) E_2^4  \right.\nn\\
&&\left.
   -\left(\frac{20125 E_6^6}{4
   E_4^6}{+}\frac{11223703 E_6^4}{120 E_4^3}{+}\frac{92285669 E_6^2}{1080}{+}\frac{1372051 E_4^3}{270}\right)
   E_2^3{+}\left(\frac{154401743 E_6^5}{360 E_4^4}{+}\frac{576047063 E_6^3}{360 E_4}\right.\right.\nn\\
&&\left.\left.{+}\frac{328463299E_4^2
   E_6}{630} \right) E_2^2
   -\left(\frac{14652664 E_6^6}{45 E_4^5}+\frac{13723519199 E_6^4}{3600 E_4^2}+\frac{11480517509 E_4
   E_6^2}{3150}\right) E_2\right.\nn\\
&&-\left.\left(\frac{753433829 E_4^4}{3150}\right) E_2+ h_4(q)\right]
 \eea
 }
The ambiguous part is given by
\bea\label{AmbiguitiesH0}
h_2 &=& \frac{1619}{15} E_6\nn\\
h_3 &=& -\frac{140891 E_6^4}{45 E_4^3}-\frac{1206371
   E_6^2}{90}-\frac{124319 E_4^3}{63}\\
h_4 &=&    \frac{26737369 E_6^7}{540 E_4^6}+\frac{7883698699
   E_6^5}{3600 E_4^3}+\frac{21429183673
   E_6^3}{4050}+\frac{25632734639 E_4^3 E_6}{18900}\nn
\eea
which has been determined by imposing the gap conditions \eqref{Bernoulli1}.

 \subsubsection*{The conformal limit}

  The theory becomes conformal in the limit $c\to 0$  and fits into the class {\bf A} according to \eqref{Classification}.
   In this limit $\tau\to e^{\pi i/3}$. Therefore $E_2$, $E_6$ become constants\footnote{Their numerical values are $E_2\approx1.103$, $E_6\approx2.881$.} and $E_4$ vanishes. More precisely, using \eqref{formulaeH0}, we find
  \bea\label{confh0}
   u &\approx& \left(\frac{5^6}{2^93^3E_6}\right)^{\frac{1}{5}}\,a^{\tfrac65}~,\nn\\
    E_4 &\approx &  \left(  \frac{ 2^6  E_6^4 }{ 3^3 \,5^4 \,}   \right)^{\frac{1}{5}}   c \,a^{-\frac{4}{5}}~,\nn\\
    \xi  &\approx &  \left(  \frac{ 2^{11} 3^2 }{E_6 5^9 \,}   \right)^{\frac{1}{5}} c\,a^{-\frac{9}{5}}~.
 \eea
 From the above formulae we notice that while both $E_4$ and $\xi$ go to zero in the limit $c\to 0$, their ratio stays finite and goes like
 \be\label{E4xiLimit}
\frac{\xi}{E_4}\approx \frac{6}{5\,E_6\,a}\,.
 \ee
Keeping only the leading terms in \eqref{FgH0cu} and \eqref{AmbiguitiesH0}, and using \eqref{E4xiLimit}, one finds
\bea
{\cal F}_2 &\approx& -\frac{7\,E_2}{800\,a^2}~,\nn\\
{\cal F}_3 &\approx&-\frac{7\, E_2^2}{8000\,a^4}~,\nn\\
{\cal F}_4 &\approx& -\frac{161\, E_2^3}{768000\,a^6}+\frac{26737369 \,E_6}{12960000000\,a^6}~.
\eea
The above formulae reproduce the result \eqref{zu}, with
\be
\beta=1\,,\quad \delta=3\,,\quad \gamma=\frac{3}{10}\,,\quad \kappa=\frac{E_6}{E_2^3}\,,\quad x=\frac{E_2\epsilon^2}{6 a^2}\,,
\ee
and
\be
 c_0=1 \quad, \quad c_1=-\frac{26737369   }{2^8\ 3\ 5^7}. \\
\ee
Higher-genus prepotentials $\mathcal{F}_g$  can also be computed. Results for the ambiguity coefficients $c_n$ are listed in \eqref{cnH0}.
As we can see,  the growth of $c_n$ is relatively fast. It is likely that the sum over hypergeometric is divergent. However, a more detailed analysis is needed.

   \subsection{$\mathcal{H}_1$ theory}\label{Sec:H1b0}

 The SW curve for the $\mathcal{H}_1$ theory deformed by the second-order mass Casimir is
 \be
 y^2=4x^3-ux-m^2\,.
 \ee
In this case $g_2=u$ and $g_3=m^2$ leading to
 \bea
 u &=& \frac{3 E_4 m^{\frac{4}{3}}}{E_6^{\frac{2}{3}}} \qquad , \qquad
  \omega_1 =\left( {8 E_6\over 27 m^2 }\right)^{1\over 6}     \qquad , \qquad  \xi =\frac{\sqrt{\frac{3}{2}} E_6^{\frac{3}{2}}}{2m(E_4^3- E_6^2 )} \nn\\
 {\cal F}_1 &=&  \frac{1}{12}  \log \left( m^6 { E_4^3-E_6^2 \over E_6^3} \right)  \quad , \quad \Delta=16(u^3-27 m^4) \label{formulaeH1}
 \eea
 Here the holomorphic ambiguity takes the form of the second expression in (\ref{iminmax}).
 Solving recursively the holomorphic anomaly equation (\ref{HAE}),  one finds the first few terms
 {\small
  \bea\label{FgH1cu}
&&  {\cal F}_2 =  {  \xi^2\over 24 \, 12^2}  \left[ \frac{5 }{3}E_2^3+\frac{3 E_4^2 }{E_6}E_2^2+\left(-\frac{9 E_4^4}{E_6^2}-46 E_4\right) E_2   + h_2(q) \right] \\
 &&  {\cal F}_3 = { \xi^4\over 24 \, 12^4}  \left[ \frac{5 }{6}E_2^6+\frac{10 E_4^2 }{E_6}E_2^5+\left(\frac{63 E_4^4}{2 E_6^2}+10 E_4\right) E_2^4+\left(\frac{9
   E_4^6}{E_6^3}-\frac{461 E_4^3}{3 E_6}-\frac{310 E_6}{9}\right) E_2^3     \right.  \nn\\
   && \quad \left.  +\left(-\frac{27 E_4^8}{E_6^4}-\frac{7068
   E_4^5}{5 E_6^2}-\frac{6871 E_4^2}{6}\right) E_2^2+\left(\frac{8289 E_4^7}{5 E_6^3}+\frac{9425 E_4^4}{E_6}+\frac{6716
   E_6 E_4}{3}\right) E_2   +h_3(q) \right]~.  \nn
    \eea
 }
The ambiguous part is given by
\bea\label{AmbiguitiesH1}
h_2 &=& \frac{351 E_4^3}{5 E_6}+\frac{566 E_6}{15} \nn\\
h_3 &=& -\frac{1112 E_4^9}{9 E_6^4}-\frac{4842049 E_4^6}{630 E_6^2}-\frac{3186886 E_4^3}{315}-\frac{12220 E_6^2}{21}
\eea
 which has been determined by imposing the gap conditions \eqref{Bernoulli1}.

   \subsubsection*{The conformal limit}

 The theory becomes conformal in the limit $m\to 0$  and fits into the class {\bf B} according to \eqref{Classification}.
   In this limit $\tau\to  {\rm i }$. Therefore $E_2$, $E_4$ become constants\footnote{Their numerical values are $E_2\approx0.955$, $E_4\approx1.456$.} and $E_6$ vanishes. More precisely, using \eqref{formulaeH1}, we find
  \bea
   u &\approx& \left(\frac{3^5}{2^{10} E_4}\right)^{\frac{1}{3}}\,a^{\tfrac43}\nn\\
    E_6 &\approx &    \frac{ 32 m^2   E_4^2 }{ 3  \, a^2}  \nn\\
    \xi  &\approx &  \frac{  64 m^2 }{ 3 a^3 \,}
 \eea
 From the above formulae we notice that while both $E_6$ and $\xi$ go to zero in the limit $m\to 0$, their ratio stays finite and goes like
 \be\label{E6xiLimit}
\frac{\xi}{E_6}\approx \frac{2}{ a \, E_4^2 }\,.
 \ee
Keeping only the leading terms in \eqref{FgH1cu} and \eqref{AmbiguitiesH1}, and using \eqref{E6xiLimit}, one finds
\bea
{\cal F}_2 &\approx& -\frac{\,E_2}{96\,a^2}\nn\\
{\cal F}_3 &\approx&-\frac{243\, E_2^2+1112 E_4}{279936\,a^4}\nn\\
\eea
The above formulae reproduce the result \eqref{zu}, with
\be
\beta=1\,,\quad \delta=2\,,\quad \gamma=\frac{1}{2}\,,\quad \kappa=\frac{E_4}{E_2^2}\,,\quad x=\frac{E_2\epsilon^2}{6 a^2}\,,
\ee
and
    \be
    c_0=1 \quad, \quad c_1=-\frac{139 }{972}  \,.
    \ee
    Higher-genus prepotentials $\mathcal{F}_g$  can also be computed. Results for the ambiguity coefficients $c_n$ are listed in \eqref{cnH1}.

      \subsection{$\mathcal{H}_2$ theory}

  The SW curve for the $\mathcal{H}_2$ theory deformed by the second-order mass Casimir is
 \be
 y^2=4x^3-m^2x-u^2\,.
 \ee
In this case $g_2=m^2$ and $g_3=u^2$ leading to
 \bea
 u &=&\frac{\sqrt{E_6} m^{\frac{3}{2}}}{3^{\frac{3}{4}} E_4^{\frac{3}{4}}}  \qquad , \qquad    \omega_1 =\left( { 4 E_4 \over 3 m^2 } \right)^{1\over 4}   \qquad , \qquad \xi =\frac{3 \sqrt{2} E_4^{\frac{3}{2}} \sqrt{E_6}}{\left(E_6^2-E_4^3\right) m}  \nn\\
 {\cal F}_1 &=& \frac{1}{12} \log \left(\frac{\left(E_4^3-E_6^2\right) m^9}{E_4^{\frac{9}{2}}}\right)  +{\rm const}  \quad , \quad \Delta=16(m^6-27 u^4)
 \label{formulaeH2}
 \eea
 Here the holomorphic ambiguity takes again the form of the first expression in (\ref{iminmax}).
 Solving recursively (\ref{HAE})  one finds the first few terms
 {\small
  \bea\label{FgH2cu}
&&  {\cal F}_2 =  {  \xi^2\over 24 \, 12^2}  \left[ \frac{5 }{3}E_2^3+\frac{3 E_4^2 }{E_6}E_2^2+\left(-\frac{3 E_6^2}{E_4^2}-52 E_4\right) E_2  + h_2(q) \right] \\
 &&  {\cal F}_3 = { \xi^4\over 24 \, 12^4}  \left[ \frac{5 }{6}E_2^6{+}\frac{5 \left(5 E_4^3{+}7 E_6^2\right) }{6 E_4 E_6}E_2^5{+}\left(\frac{2 E_4^4}{E_6^2}{+}\frac{185
   E_4}{6}{+}\frac{26 E_6^2}{3 E_4^2}\right) E_2^4{-} \left(\frac{251 E_4^3}{6 E_6}{+}\frac{1207 E_6}{9}{+}\frac{19 E_6^3}{6
   E_4^3}\right) E_2^3   \right.  \nn\\
   && \left. {-}\left(\frac{153 E_4^5}{2 E_6^2}{+}\frac{9167 E_4^2}{6}{+}\frac{29353 E_6^2}{30 E_4}{+}\frac{3
   E_6^4}{E_4^4}\right) E_2^2{+}\left(\frac{2343 E_4^4}{E_6}{+}\frac{277747 E_6 E_4}{30}{+}\frac{51607 E_6^3}{30
   E_4^2}\right) E_2    {+}h_3(q) \right]  \nn\\
 &&  {\cal F}_4= { \xi^6\over 24 \, 12^6}  \left[
 \frac{1105 E_2^9}{1296}+\left(\frac{985 E_4^2}{144 E_6}+\frac{805 E_6}{72 E_4}\right)
   E_2^8+\left(\frac{445 E_4^4}{36 E_6^2}+\frac{8135 E_4}{72}+\frac{3781 E_6^2}{72
   E_4^2}\right) E_2^7
    \right.  \nn\\
   && \left.
   +\left(\frac{11 E_4^6}{4 E_6^3}+\frac{54395 E_4^3}{216
   E_6}+\frac{117511 E_6}{216}+\frac{10921 E_6^3}{108 E_4^3}\right) E_2^6+\left(\frac{59
   E_4^5}{2 E_6^2}-\frac{8509 E_4^2}{48}-\frac{4097 E_6^2}{36 E_4}+\frac{13583 E_6^4}{240
   E_4^4}\right) E_2^5
    \right.  \nn\\
   && \left.
   -\left(\frac{99 E_4^7}{2 E_6^3}+\frac{1176895 E_4^4}{144
   E_6}+\frac{9441703 E_6 E_4}{360}+\frac{1150283 E_6^3}{144 E_4^2}+\frac{577 E_6^5}{24
   E_4^5}\right) E_2^4
    \right.  \nn\\
   && \left.
   -\left(\frac{7273 E_4^6}{E_6^2}{+}\frac{775503
   E_4^3}{10}{+}\frac{97040443 E_6^2}{1080}{+}\frac{15561623 E_6^4}{1080 E_4^3}
   -\frac{9
   E_6^6}{E_4^6}\right) E_2^3
    \right.  \nn\\
   && \left.
   +\left(\frac{13905 E_4^8}{4 E_6^3}+\frac{6407761 E_4^5}{18
   E_6}+\frac{512201711}{360} E_6 E_4^2+\frac{29391479 E_6^3}{40 E_4}+\frac{42036497
   E_6^5}{1260 E_4^4}\right) E_2^2
    \right.  \nn\\
   && \left.
   -\left(\frac{160687 E_4^7}{E_6^2}+\frac{10391931
   E_4^4}{4}+\frac{385527557}{90} E_6^2 E_4+\frac{8137162319 E_6^4}{8400
   E_4^2}+\frac{3300704 E_6^6}{315 E_4^5}\right) E_2
  +h_4(q) \right]
 \eea
 }
The ambiguous part is given by
\bea\label{AmbiguitiesH2}
h_2 &=& \frac{147 E_4^3}{5 E_6}+\frac{1178 E_6}{15} \nn\\
h_3 &=&-\frac{3529 E_4^6}{14 E_6^2}-\frac{1038589 E_4^3}{126}-\frac{6008447 E_6^2}{630}-\frac{150032 E_6^4}{315 E_4^3} \\
h_4 &=& \frac{63691 E_4^9}{10 E_6^3}+\frac{25347539 E_4^6}{30 E_6}+\frac{132133663}{30} E_6 E_4^3+\frac{150291551071
   E_6^3}{45360}\nn\\&& +\frac{11994210803 E_6^5}{37800 E_4^3}+\frac{12428 E_6^7}{27 E_4^6} \nn
\eea
  which has been determined by imposing the gap conditions  \eqref{Bernoulli1}.

  \subsubsection*{The conformal limit}

 The theory becomes conformal in the limit $m\to 0$  and fits into the class {\bf A} according to \eqref{Classification}.
   In this limit $\tau\to e^{\pi i/3}$. Therefore $E_2$, $E_6$ become constants\footnote{Their values are clearly the same as in the $\mathcal{H}_0$ theory.} and $E_4$ vanishes. More precisely, using \eqref{formulaeH2}, we find
  \bea
   u &\approx&  \left(\frac{8}{27 E_6}\right)^{\tfrac14}\,a^{\tfrac32}\nn\\
    E_4 &\approx &   \frac{  m^2 E_6  }{ 2 a^2 }  \nn\\
    \xi  &\approx &   \frac{ 3 m^2 }{ 2 a^3  \,}  \label{confh2}
 \eea
 From the above formulae we notice that while both $E_4$ and $\xi$ go to zero in the limit $m\to 0$, their ratio stays finite and goes like
 \be\label{E4xiLimit2}
\frac{\xi}{E_4}\approx  \frac{3}{E_6\,a}\,.
 \ee
Keeping only the leading terms in \eqref{FgH2cu} and \eqref{AmbiguitiesH2}, and using \eqref{E4xiLimit2}, one finds
\bea
{\cal F}_2 &\approx& -\frac{E_2}{128\,a^2}\nn\\
{\cal F}_3 &\approx&-\frac{E_2^2}{2048\,a^4}\nn\\
{\cal F}_4 &\approx&- \frac{243\, E_2^3-12428\, E_6 }{2654208 \,a^6}
\eea
The above formulae reproduce the result \eqref{zu}, with
\be
\beta=1\,,\quad \delta=3\,,\quad \gamma=\frac{3}{4}\,,\quad \kappa=\frac{E_6}{E_2^3}\,,\quad x=\frac{E_2\epsilon^2}{6 a^2}\,,
\ee
and
    \be\label{csH2}
    c_0=1 \quad, \quad c_1= -\frac{3107}{3072} \,.
   \ee
 Higher-genus prepotentials $\mathcal{F}_g$  can also be computed. Results for the ambiguity coefficients $c_n$ are listed in \eqref{cnH2}.

  \subsection{E$_6$ theory}

The SW curve for the E$_6$ theory deformed by the eighth-order mass Casimir is
 \be
 y^2=4x^3-m^8x-u^4\,.
 \ee
In this case $g_2=m^8$ and $g_3=u^4$ leading to
 \bea
 u &=&\frac{E_6^{\frac{1}{4}} m^3}{3^{\frac{3}{8}} E_4^{\frac{3}{8}}}  \qquad , \qquad    \omega_1 =\left( { 4 E_4 \over 3 m^8 } \right)^{1\over 4}   \qquad , \qquad \xi =\frac{2 \sqrt{2}\, 3^{\frac{5}{8}} E_4^{\frac{9}{8}} E_6^{\frac{3}{4}}}{\left(E_6^2-E_4^3\right) m}  \nn\\
 {\cal F}_1 &=& \frac{1}{12} \log \left(\frac{\left(E_4^3-E_6^2\right) m^{36}}{E_4^{\frac{9}{2}}}\right)  +{\rm const}  \quad , \quad \Delta=16(m^{24}-27 u^8)
 \label{formulaeE6}
 \eea
 {\small
 Here the holomorphic ambiguity takes again the form of the first expression in (\ref{iminmax}).
 Solving recursively (\ref{HAE})  one finds the first few terms
  \bea\label{FgE6cu}
&&  {\cal F}_2 =  {  \xi^2\over 24 \, 12^2}  \left[ \frac{5 E_2^3}{3}+\left(\frac{9 E_4^2}{2 E_6}-\frac{3 E_6}{2 E_4}\right)
   E_2^2+\left(\frac{6 E_6^2}{E_4^2}-61 E_4\right) E_2 + h_2(q) \right] \\
 &&  {\cal F}_3 = { \xi^4\over 24 \, 12^4}  \left[ \frac{5 E_2^6}{6}+\frac{5 \left(5 E_4^3+3 E_6^2\right) E_2^5}{4 E_4 E_6}+\left(\frac{39
   E_4^4}{4 E_6^2}+\frac{115 E_4}{4}+\frac{3 E_6^2}{E_4^2}\right) E_2^4 \right.  \nn\\
   && \left.+\frac{\left(81
   E_4^6-4869 E_6^2 E_4^3-8165 E_6^4+\frac{57 E_6^6}{E_4^3}\right) E_2^3}{72
   E_6^3}+\left(-\frac{1566 E_4^5}{5 E_6^2}-\frac{107869 E_4^2}{60}-\frac{474
   E_6^2}{E_4}-\frac{3 E_6^4}{4 E_4^4}\right) E_2^2 \right.  \nn\\
  && \left. +\frac{\left(3969 E_4^9+562005 E_6^2
   E_4^6+957017 E_6^4 E_4^3+75585 E_6^6\right) E_2}{120 E_4^2 E_6^3}+h_3(q)\right]\nn\\
 &&  {\cal F}_4= { \xi^6\over 24 \, 12^6}  \left[
\frac{1105 E_2^9}{1296}+\left(\frac{985 E_4^2}{96 E_6}+\frac{745 E_6}{96 E_4}\right)
   E_2^8+\left(\frac{303 E_4^4}{8 E_6^2}+\frac{8399 E_4}{72}+\frac{70 E_6^2}{3
   E_4^2}\right) E_2^7\right.  \nn\\
   && \left.
   +\frac{\left(19143 E_4^9 +198654 E_6^2 E_4^6+159643
   E_6^4E_4^3+11244 E_6^6\right) E_2^6}{432 E_6^3E_4^3}   \right.  \nn\\
   && \left.
  +\frac{\left(27459
   E_4^{12}+638676 E_6^2 E_4^9-734194 E_6^4 E_4^6-547740 E_6^6 E_4^3+25455 E_6^8\right)
   E_2^5}{2880 E_4^4 E_6^4}
    \right.  \nn\\
   && \left.
   -\frac{\left(2599047 E_4^{12}+47027130 E_6^2 E_4^9+64395032
   E_6^4 E_4^6+8268330 E_6^6 E_4^3-555 E_6^8\right) E_2^4}{2880 E_4^5
   E_6^3}     \right.  \nn\\
   && \left.
   -\frac{\left(1954449 E_4^{15}+135004860
   E_4^{12}E_6^2+452446390 E_4^9 E_6^4+215240996 E_4^6E_6^6+12237645 E_6^8E_4^3-540
   E_6^{10}\right) E_2^3}{4320 E_6^4 E_4^6}
    \right.  \nn\\
   && \left.
   +\frac{\left(484499421 E_4^{12}+8105065554 E_6^2
   E_4^9+13840109482 E_6^4 E_4^6+3239239130 E_6^6 E_4^3+39065765 E_6^8\right)
   E_2^2}{10080 E_4^4 E_6^3}\right.\nn\\
   &&\left.{-}\frac{\left(482143347 E_4^{15}+70093473948 E_6^2
   E_4^{12}+394187429578 E_6^4 E_4^9+312678218260 E_6^6 E_4^6\right) E_2}{100800 E_4^5 E_6^4}\right.\nn\\&&\left.{-} \frac{\left(31095995615 E_6^8
   E_4^3+29687000 E_6^{10}\right)E_2}{100800 E_4^5 E_6^4}
  +h_4(q) \right]
 \eea
 }
The ambiguous part is given by
\bea\label{AmbiguitiesE6}
h_2 &=& \frac{441 E_4^3}{10 E_6}+\frac{383 E_6}{6} \nn\\
h_3 &=&-\frac{21909 E_4^6}{20 E_6^2}-\frac{3426562 E_4^3}{315}-\frac{4063991 E_6^2}{630}-\frac{21205 E_6^4}{252 E_4^3} \\
h_4 &=& \frac{150204789 E_4^9}{1600 E_6^3}+\frac{2879128369 E_4^6}{1400 E_6}+\frac{731025235537 E_6 E_4^3}{151200}\nn\\ &&+\frac{166127444801
   E_6^3}{90720}+\frac{22522691 E_6^5}{320 E_4^3}+\frac{8 E_6^7}{27 E_4^6} \nn
\eea
  which has been determined by imposing the gap conditions  \eqref{Bernoulli1}.

  \subsubsection*{The conformal limit}

 The theory becomes conformal in the limit $m\to 0$  and fits into the class {\bf A} according to \eqref{Classification}.
   In this limit $\tau\to e^{\pi i/3}$. Therefore $E_2$, $E_6$ become constants\footnote{Their values are clearly the same as in the $\mathcal{H}_0$ and $\mathcal{H}_2$ theories.} and $E_4$ vanishes. More precisely, using \eqref{formulaeE6}, we find
  \bea
   u &\approx&  \frac{a^3}{6\sqrt{6} E_6^{\tfrac12}}\nn\\
    E_4 &\approx &   \frac{2^4 \, 3^3 E_6^2\, m^8   }{ a^8 }  \nn\\
    \xi  &\approx &   \frac{ 2^6 \, 3^4 E_6 \,m^8 }{ a^9  \,}  \label{confE6}
 \eea
 From the above formulae we notice that while both $E_4$ and $\xi$ go to zero in the limit $m\to 0$, their ratio stays finite and goes like
 \be\label{E4xiLimitE6}
\frac{\xi}{E_4}\approx  \frac{12}{E_6\,a}\,.
 \ee
Keeping only the leading terms in \eqref{FgE6cu} and \eqref{AmbiguitiesE6}, and using \eqref{E4xiLimitE6}, one finds
\bea
{\cal F}_2 &\approx& \frac{E_2}{4a^2}\nn\\
{\cal F}_3 &\approx&-\frac{E_2^2}{32a^4}\nn\\
{\cal F}_4 &\approx&\left(\frac{E_2^3}{192}+\frac{E_6 }{81}\right)\frac{1}{a^6}
\eea
The above formulae reproduce the result \eqref{zu}, with
\be
\beta=1\,,\quad \delta=3\,,\quad \gamma=3\,,\quad \kappa=\frac{E_6}{E_2^3}\,,\quad x=\frac{E_2\epsilon^2}{6 a^2}\,,
\ee
and
    \be\label{csE6}
    c_0=1 \quad, \quad c_1= -\frac{8}{3} \,.
   \ee
 Higher-genus prepotentials $\mathcal{F}_g$  can also be computed in exactly the same manner as in the previous examples.

 \subsection{E$_7$ theory}

The SW curve for the E$_7$ theory deformed by the eighteenth-order mass Casimir is
\be
y^2=4x^3-u^3x-m^{18}\,.
\ee
In this case $g_2=u^3$ and $g_3=m^{18}$ leading to

\bea
u &=&\frac{3^\frac{1}{3} m^4 E_4^\frac{1}{3}}{E_6^\frac{2}{9}}  \qquad , \qquad
 \omega_1 =\sqrt{\frac{2}{3}}\left( {  E_6^\frac{1}{6} \over  m^3 } \right)   \qquad , \qquad \xi =\frac{3^\frac{13}{6}E_4^\frac{2}{3}E_6^\frac{19}{18}}{2^\frac{3}{2}m(E_4^3-E_6^2)}  \nn\\
{\cal F}_1 &=& \frac{1}{12} \log \left(\frac{(E_4^3-E_6^2) m^{54}}{E_6^3}\right)  +{\rm const}  \quad , \quad \Delta=16(u^{9}-27 m^{36})
\label{formulaeE7}
\eea
Here the holomorphic ambiguity takes again the form of the second expression in (\ref{iminmax}).
Solving recursively (\ref{HAE})  one finds the first few terms
{\small
	\bea\label{FgE7cu}
	&&  {\cal F}_2 =  {  \xi^2\over 24 \, 12^2}  \left[ \frac{5 E_2^3}{3}+\left(\frac{E_4^2}{3 E_6}+\frac{8 E_6}{3 E_4}\right)E_2^2+\left(\frac{7 E_4^4}{E_6^2}-62 E_4\right) E_2
	+ h_2(q) \right] \\
	&&{\cal F}_3 = { \xi^4\over 24 \, 12^4} \left[\frac{5  E_2^6}{6}+\frac{10  E_2^5 (17  E_4^3+10  E_6^2)}{27  E_4  E_6}+ E_2^4 (\frac{191  E_4^4}{18  E_6^2}+\frac{80  E_6^2}{27  E_4^2}+\frac{754  E_4}{27})\right.\nn\\
	 &&\left.
	 +\frac{ E_2^3 (229  E_4^6-7487  E_4^3  E_6^2-7250  E_6^4)}{81E_6^3}-\frac{ E_2^2 (210  E_4^9+142680  E_4^6  E_6^2+577955  E_4^3  E_6^4+101728  E_6^6)}{270  E_4  E_6^4}\right.\nn\\
	 &&\left. -\frac{ E_2 (6905  E_4^9+128405  E_4^6
		 E_6^2+59572  E_4^3  E_6^4-2624  E_6^6)}{135  E_4^2  E_6^3}+h_3(q)\right]
	\eea
}
The ambiguous part is given by
\bea\label{AmbiguitiesE7}
h_2 &=& -\frac{191 E_4^3}{15 E_6}+\frac{82 E_6}{15} \nn\\
h_3 &=&-\frac{302161 E_6^4}{5184 E_4^3}-\frac{2294657 E_4^3}{2592}+\frac{230203 E_6^2}{3240}-\frac{1544857 E_4^9}{25920 E_6^4} \\
\eea
which has been determined by imposing the gap conditions  \eqref{Bernoulli1}. % In this case we give the result up to $g=3$ since ambiguity enters the final result already at  this order.

\subsubsection*{The conformal limit}

The theory becomes conformal in the limit $m\to 0$  and fits into the class {\bf B} according to \eqref{Classification}.
In this limit $\tau\to i$. Therefore $E_2$, $E_4$ become constant\footnote{Their values are clearly the same as in the $\mathcal{H}_1$ theory.} and $E_6$ vanishes. More precisely, using \eqref{formulaeE7}, we find
\bea
u &\approx& \frac{3\,a^4}{2^{10}\,E_4} \nn\\
E_6 &\approx & \frac{2^{45}\, E_4^6\, m^{18}   }{ 3^3\, a^{18} } \nn\\
\xi  &\approx &  \frac{2^{46}\, E_4^4\, m^{18}   }{ 3\, a^{19} }  \label{confE7}
\eea
From the above formulae we notice that while both $E_6$ and $\xi$ go to zero in the limit $m\to 0$, their ratio stays finite and goes like
\be\label{E4xiLimitE7}
\frac{\xi}{E_6}\approx  \frac{18}{E_4^2\,a}\,.
\ee
Keeping only the leading terms in \eqref{FgE7cu} and \eqref{AmbiguitiesE7}, and using \eqref{E4xiLimitE7}, one finds
\bea
{\cal F}_2 &\approx& \frac{21E_2}{32a^2}\nn\\
{\cal F}_3 &\approx&-\left(\frac{21 E_2^2}{128}+\frac{1544857 E_4}{330598817280}\right)\frac{1}{a^4}
\eea
The above formulae reproduce the result \eqref{zu}, with
\be
\beta=1\,,\quad \delta=2\,,\quad \gamma=\frac{9}{2}\,,\quad \kappa=\frac{E_4}{E_2^2}\,,\quad x=\frac{E_2\epsilon^2}{6 a^2}\,,
\ee
and
\be\label{csE7}
c_0=1 \quad, \quad c_1= -\frac{4634571}{10240} \,.
\ee
Higher-genus prepotentials $\mathcal{F}_g$  can also be computed in exactly the same manner as in the previous examples.

\subsection{E$_8$ theory}

The SW curve for the E$_8$ theory deformed by the twentyth-order mass Casimir is
 \be
 y^2=4x^3-m^{20}x-u^5\,.
 \ee
In this case $g_2=m^{20}$ and $g_3=u^5$ leading to
 \bea
 u &=&\frac{E_6^{\frac{1}{5}} m^6}{3^{\frac{3}{10}} E_4^{\frac{3}{10}}}  \qquad , \qquad    \omega_1 =\left( { \sqrt2 E_4^{\frac{1}{4}} \over 3^{\frac{1}{4}} m^5 } \right)   \qquad , \qquad
 \xi =\frac{5\, 3^\frac{11}{20} E_4^{\frac{21}{20}} E_6^{\frac{4}{5}}}{\sqrt2 m\left(E_6^2-E_4^3\right)}  \nn\\
 {\cal F}_1 &=& \frac{1}{12} \log \left(\frac{\left(E_4^3-E_6^2\right) m^{90}}{E_4^{\frac{9}{2}}}\right)  +{\rm const}  \quad , \quad \Delta=16(m^{60}-27 u^{10})
 \label{formulaeE8}
 \eea
 Here the holomorphic ambiguity takes again the form of the first expression in (\ref{iminmax}).
 Solving recursively (\ref{HAE})  one finds the first few terms
 {\small
  \bea\label{FgE8cu}
&&  {\cal F}_2 =  {  \xi^2\over 24 \, 12^2}  \left[ \frac{5 E_2^3}{3}+\left(\frac{24 E_4^2}{5 E_6}-\frac{9 E_6}{5 E_4}\right)
   E_2^2+\left(\frac{39 E_6^2}{5 E_4^2}-\frac{314 E_4}{5}\right) E_2 + h_2(q) \right] \nn \\
   &&  {\cal F}_3 = { \xi^4\over 24 \, 12^4}  \left[ \frac{5 E_2^6}{6}+\frac{10 \left(2 E_4^3+ E_6^2\right) E_2^5}{3 E_4 E_6}+
 \left(\frac{
   1776 E_4^4}{150 E_6^2}+\frac{4088 E_4}{150}+\frac{361 E_6^2}{150 E_4^2}\right) E_2^4 \right.  \nn\\
   && \left.
   +\frac{\left(2592  E_4^9-85236 E_6^2 E_4^6-119429 E_4^3 E_6^4+573 E_6^6\right) E_2^3}{1125
   E_6^3 E_4^3}
   -\left(-\frac{307080 E_4^5}{750 E_6^2}+\frac{ 1603513 E_4^2}{750}-\frac{468 E_6^4}{750 E_4^4}\right) E_2^2  \right.  \nn\\
   && \left. +\left(\frac{373864 E_6^2}{750 E_4}\right) E_2^2 -\frac{\left(-13392 E_4^9+705816 E_6^2
   E_4^6+1812719 E_6^4 E_4^3+165107 E_6^6\right) E_2}{1875 E_4^2 E_6^3}+h_3(q)\right]
   \eea
\bea\label{FgE8cu2}
 &&  {\cal F}_4= { \xi^6\over 24 \, 12^6}  \left[
\frac{1105 E_2^9}{1296}+\left(\frac{197 E_4^2}{18 E_6}+\frac{745 E_6}{96 E_4}\right)
   E_2^8
   +\left(\frac{80144 E_4^4}{1800 E_6^2}+\frac{205727 E_4}{1800}+\frac{34279 E_6^2}{1800
   E_4^2}\right) E_2^7\right.  \nn\\
   && \left.
   +\frac{\left(1660608 E_4^9 +13257736 E_6^2 E_4^6+495277
   E_6^4E_4^3+11244 E_6^6\right) E_2^6}{27000 E_6^3E_4^3}   \right.  \nn\\
   &&\left.+\frac{\left(8688384 E_4^{12}+117129744 E_6^2 E_4^9-201837711
   E_6^4 E_4^6-104795596 E_6^6 E_4^3+2173929 E_6^8\right) E_2^5}{450000 E_4^4
   E_6^4}     \right.  \nn\\
&&\left.
-\frac{\left(120661632 E_4^{12}+1894444360 E_6^2 E_4^9+2294021113 E_6^4 E_4^6+241621016 E_6^6 E_4^3-1371 E_6^8\right)
   E_2^4}{2880 E_4^4 E_6^4}+\frac{39 E_6^6\,E_2^3}{500 E_4^6}
    \right. \nn\\
    && \left.
   -\frac{\left(657891072 E_4^{12}+43610234472
   E_4^{9}E_6^2+155402608452 E_4^6 E_6^4+70330783117 E_4^3E_6^6+2529429287 E_6^8
   \right) E_2^3}{675000 E_6^4 E_4^3}
    \right.  \nn\\&& \left.
   -\frac{\left(211652806254 E_4^{9}+3413887736674 E_6^2
   E_4^6+5487101754954 E_6^4 E_4^3+1215593719179 E_6^6
   \right) E_2^2}{42525000 E_4 E_6^3}\right.\nn\\
   &&+\left.+\frac{2142744527 E_6^5 E_2^2}{6075000 E_4^4}+\frac{\left(2590531017336 E_4^{10}+248754470552736 E_6^2
   E_4^{7}+1207090705967416 E_6^4 E_4^4\right) E_2}{425250000 E_6^4}\right.\nn\\
   &&\left.+\frac{35035331393449 E_6^2 E_4 E_2}{8859375}+ \frac{\left(70923207769701 E_6^8
   E_4^3+31893238160 E_6^{10}\right)E_2}{425250000 E_4^5 E_6^4}
  +h_4(q) \right]
 \eea
 }
The ambiguous part is given by
\bea\label{AmbiguitiesE8}
h_2 &=& \frac{124 E_4^3}{25 E_6}-\frac{917 E_6}{75} \nn\\
h_3 &=&\frac{21308148037 E_4^3}{2835000}+\frac{11593299743 E_6^2}{2835000}+\frac{56925211 E_6^4}{1215000 E_4^3}+\frac{7183179683 E_4^6}{8505000 E_6^2}\nn \\
h_4 &=& -\frac{392331535221859 E_6^3}{273375000}-\frac{460255802444281 E_6 E_4^3}{1913625000}-\frac{8109292812051391 E_4^6}{3827250000 E_6}\nn\\
&&-\frac{460255802444281 E_4^9}{3827250000 E_6^3}-\frac{12356384824399 E_6^5}{273375000 E_4^3}+\frac{4482151319 E_6^7}{109350000 E_4^6}
\eea
  which has been determined by imposing the gap conditions  \eqref{Bernoulli1}.

  \subsubsection*{The conformal limit}

 The theory becomes conformal in the limit $m\to 0$  and fits into the class {\bf A} according to \eqref{Classification}.
   In this limit $\tau\to e^{\pi i/3}$. Therefore $E_2$, $E_6$ become constants and $E_4$ vanishes. More precisely, using \eqref{formulaeE8}, we find
  \bea
   u &\approx& \frac{a^6}{2^9\,3^3\,E_6}  \nn\\
    E_4 &\approx &  \frac{3^9\, 2^{30}\, E_6^4\, m^{20}   }{  a^{20} } \nn\\
    \xi  &\approx &  \frac{5\,2^{31}\,3^{10}\, E_6^3\, m^{20}   }{  a^{21} } \label{confE8}
 \eea
 From the above formulae we notice that while both $E_4$ and $\xi$ go to zero in the limit $m\to 0$, their ratio stays finite and goes like
 \be\label{E4xiLimitE8}
\frac{\xi}{E_4}\approx  \frac{30}{E_6\,a}\,.
 \ee
Keeping only the leading terms in \eqref{FgE8cu}, \eqref{FgE8cu2} and \eqref{AmbiguitiesE8}, and using \eqref{E4xiLimitE8}, one finds
\bea
{\cal F}_2 &\approx& \frac{65 E_2}{32 a^2}\nn\\
{\cal F}_3 &\approx&-\frac{65 E_2^2}{64 a^4}\nn\\
{\cal F}_4 &\approx& \left(\frac{1625 E_2^3}{2048}+\frac{22410756595 E_6 }{53747712}\right)\frac{1}{a^6}
\eea
The above formulae reproduce the result \eqref{zu}, with
\be
\beta=1\,,\quad \delta=3\,,\quad \gamma=\frac{15}{2}\,,\quad \kappa=\frac{E_6}{E_2^3}\,,\quad x=\frac{E_2\epsilon^2}{6 a^2}\,,
\ee
and
    \be\label{csE8}
    c_0=1 \quad, \quad c_1=\frac{22410756595}{248832} \,.
   \ee
 Higher-genus prepotentials $\mathcal{F}_g$  can also be computed in exactly the same manner as in the previous examples.

  \section{NS limit: $\beta=0$}\label{Sec:NS}

   In this section we consider the NS limit $\epsilon_1\to 0$, i.e. $\beta\to 0$ \cite{Nekrasov:2009rc}. In this limit the prepotential takes
  the form
   \be\label{Fge2} \mathcal{F} =  \sum_{g=0} \epsilon_2^{2g} {\cal F}_g
   \ee
   where
   \be
   {\cal F}_g=\mathcal{F}_{0,g}=\lim_{\beta \to 0} \beta^{2g} \, {\cal F}_g(\beta)
   \ee
    The holomorphic anomaly \eqref{hafh} for $\widehat{\cal F}={\cal F}-{\cal F}_0$ becomes
\be   \label{nsfh}
\partial_{E_2} \widehat{\cal F}=\frac{1}{24}  \left(\partial_a  \widehat{\cal F} \right)^2  \,
\ee
 or equivalently, using \eqref{Fge2},
  \be
  \label{HAEb0}
 \partial_{E_2} \widehat{\cal F} _{g} =\frac{1}{24} \sum_{g'=1}^{g-1} \partial_a \widehat{\cal F} _{g'} \partial_a \widehat{\cal F} _{g-g'}
 \ee
This equation allows to compute all $\widehat{\cal F} _g$ terms recursively starting from $\mathcal{F}_1$, given in \eqref{F1general}. Sending $\beta\to0$ and using \eqref{Discriminantg2g3}, we get
   \be
   \widehat{\cal F} _{1}=\frac{1}{24} \log{ 1024(E_4^3-E_6^2) \over 27  \omega_1^{12}} \underset{m\to 0}{\approx} \tilde \gamma \log  \left({a\over {\epsilon_2}}\right)+{\rm const}  \label{f1b0}
   \ee
   with
 \be
  \tilde \gamma=\lim_{\beta\to 0}\beta^2 \gamma={d-1\over 2}\,.
\ee

Equation \eqref{HAEb0} has been extensively studied in the context of the WKB expansion for a certain class of quantum mechanical operators corresponding to AD theories with some deformations \cite{Codesido:2017dns,Codesido:2017jwp,CodesidoSanchez:2018vor,Gu:2022fss,Fischbach:2018yiu}.
See also \cite{Gaiotto:2014bza,Ito:2017ypt,Grassi:2018spf,Hollands:2019wbr,Hollands:2021itj} for other works relating WKB and non-lagrangian theories.

Here we are interested in the conformal limit where such deformationss are turned off. From the point of view of quantum mechanics this corresponds to having a potential with a single term of the form $V(x)=x^n$, $n\geq3$.
Parallel to \eqref{zansatz} we  make the following Ansatz to capture the conformal limit \footnote{Recall that $\mathcal{F}$ in this paper is defined up to a multiplicative constant. }
\be \label{zansatz3}\epsilon_1^{-2}\mathcal{F}={\tilde\gamma \over 2}\log \left(E_2\right)+\sum_{n\geq 0}\left({E_{2\delta}\over E_2^\delta}\right)^n f_n(x)\,, \qquad x=\frac{E_2 \epsilon_2^2}{6 a^2}~.
\ee
Equation \eqref{nsfh} can then be solved order by order in $E_{2\delta}$. At zero order we have\footnote{We recall that in this conformal limit $a$ and $E_2$ can be treated as independent variables.}
\be\tilde\gamma -2 x^3 f_0'(x)^2+2 x f_0'(x) =0\ee
where the boundary conditions are chosen such that the solution does not have negative power-like behavior as $x\to 0$. This gives
\be\label{f0h0ns} f_0(x)=\frac{\sqrt{2 \tilde \gamma  x+1}- x \tilde\gamma  \log \left(\frac{\tilde\gamma  x+\sqrt{2 \tilde\gamma  x+1}+1}{\tilde\gamma  x}\right)-1}{2 x}~.\ee
At the first order in $E_{2\delta}$, \eqref{nsfh} gives
\be  x \left(2 x^2 f'_0(x)-1\right) f'_1(x)+\delta f_1(x)=0~.\ee
Using \eqref{f0h0ns} we obtain
\be \label{f1h0ns} f_1(x)=c_1\left(\frac{\tilde\gamma  x-\sqrt{2\tilde \gamma  x+1}+1}{\tilde\gamma  x}\right)^{\delta }~,\ee
where $c_1$ is the integration constant.  Likewise, at second order we find
\be f_2(x)= \left(\frac{\gamma  x-\sqrt{2 \gamma  x+1}+1}{\gamma  x}\right)^{2 \delta } \left(c_2-c_1^2\frac{ \delta ^2}{\gamma  \sqrt{2 \gamma  x+1}}\right)\ee
where $c_2$ is the integration constant and $c_1$ is as in \eqref{f1h0ns}.
 In principle higher-order $f_n$ terms can also be obtained in a similar manner. However, in contrast to the case of finite $\beta$, when $\beta=0$ we do not  have a general form for $f_n$. The value of the integration constants $c_n$ is fixed by the holomorphic ambiguity.

   \subsection*{The example of $\mathcal{H}_0$}
   Let us spell out some detail for the case of  $\mathcal{H}_0$. The starting point of the recursion is given in (\ref{f1b0}), where $m$ is the relevant coupling $c^{5/4}$.
   The special geometry relations are as in \eqref{formulaeH0} and we have
    $\tilde \gamma=1/10$, $\delta=3$.   The ambiguity  is of the form \eqref{iminmax} (first line)
   and the gap condition is  given in \eqref{gapns}.   Using these initial conditions  and running the equation \eqref{nsfh},     we obtain
   \be\ba \mathcal{F}_{2}=&{1\over 24~ 12^2}\xi^2\left[\frac{E_2 E_6^2}{3456 E_4^2}+h_2(q)\right] \\
   \mathcal{F}_{3}=&{1\over 24~ 12^4}\xi^4\left[-\frac{E_2^3E_6^3}{4458050224128 E_4^3}-\frac{E_2^2 \left(6 E_4^3E_6^2+5E_6^4\right)}{1486016741376 E_4^4}-\frac{E_2 \left(474 E_4^3E_6+1427E_6^3\right)}{7430083706880 E_4^2}
   +h_3(q)
   \right]\\
    \mathcal{F}_{4}=&{1\over 24~ 12^4}\xi^4\left[\frac{E_2^5 E_6^4}{739537035580145664E_4^4}+\frac{E_2^4 \left(24E_4^3 E_6^3+23 E_6^5\right)}{739537035580145664E_4^5}\right.
    \\
    &+\frac{E_2^2 \left(5688E_4^6 E_6+55638E_4^3 E_6^3+47141 E_6^5\right)}{1848842588950364160E_4^4}+\frac{E_2^3 \left(2502E_4^6 E_6^2+8011E_4^3 E_6^4+1250 E_6^6\right)}{5546527766851092480E_4^6}\\
  & \left. +\frac{E_2 \left(1572732E_4^9+95314012E_4^6 E_6^2+199451203E_4^3 E_6^4+24620960 E_6^6\right)}{129418981226525491200E_4^5}+h_4(q)~\right]\\
   \ea\ee
   with
   \be
    \xi ={  3^{7\over 4} \, E_4^{9\over 4} \over  2^{1\over 2}  c^{5\over 4} (E_6^2-E_4^3) }~,
   \ee
   and \be\ba h_2=& \frac{79 E_6}{5}&\\
   h_3=&-\frac{21983 E_6^4}{22290251120640 E_4^3}-\frac{5611 E_4^3}{8668430991360}-\frac{47731 E_6^2}{11145125560320}\\
   h_4=&\frac{8670019 E_6^7}{19412847183978823680 E_4^6}+\frac{382204771 E_6^5}{18488425889503641600 E_4^3}\\
   &+\frac{107731843 E_4^3 E_6}{8088686326657843200}+\frac{706159453 E_6^3}{13866319417127731200}~ .\\
   \ea\ee
    The results perfectly agree with those obtained in \cite{Codesido:2017dns}. The conformal limit $c\to 0$ can be computed using
    (\ref{confh0}) and \eqref{E4xiLimit}, and gives
   \be \label{c0dwn}\ba  \widehat{\mathcal{F}}~&\approx  ~\frac{\log (a)}{10}+\frac{E_2}{2400 a^2}-\frac{E_2^2}{288000 a^4}+\frac{4375 E_2^3+8670019 E_6}{90720000000 a^6}\\
   &-\frac{34680076 E_2 E_6+6125 E_2^4}{7257600000000 a^8}+\frac{26010057 E_2^2 E_6+2450 E_2^5}{145152000000000 a^{10}}\\
   &-\frac{8523712429375 E_2^3 E_6+516140625 E_2^6+261612601805031778 E_6^2}{1401079680000000000000 a^{12}}\\
   &+\frac{155131566214625 E_2^4 E_6+14661054900894611191 E_2 E_6^2+6709828125 E_2^7}{784604620800000000000000 a^{14}}
   +O\left(\frac{1}{a^{16}}\right)~.\ea\ee
   This agrees with \eqref{zansatz3} for
   \be c_1=\frac{8670019}{52500}\qquad,\qquad c_2=-\frac{1458581050220478983}{2627625000}\,.\ee

   \section{Outlook}\label{Sec:Outlook}

   This paper  exploits the holomorphic anomaly equation to compute
     the partition function of intrinsically strongly coupled SCFT's  with eight supercharges  living on a generic $\Omega$-background.  We studied one-parameter deformations of such theories allowing for an exact integration of the anomaly equation, order by order in the $\epsilon$ expansion of the  $\Omega$-deformed prepotential.  Within this framework, we observed important simplifications at the conformal point. The  $\Omega$-deformed prepotential is given by the elegant formula \eqref{zuin}  in terms of hypergeometric functions, with coefficients $c_n$ determined by the gap conditions. It would be interesting to understand whether this non-trivial re-organization of the $\epsilon$ expansion of the partition function can improve the precision in the computation of extremal correlators made in \cite{Grassi:2019txd,Bissi:2021rei}, and shed light on the analytic structure of  the exact answer. We will report on this in \cite{GFMS}.

In the NS phase of the $\Omega$ background ($\epsilon_1=0$), we show that \eqref{zuin} undergo a non-trivial re-organization in which the hypergeometric functions become simpler functions, see e.g~\eqref{f0h0ns} and \eqref{f1h0ns}. This results are relevant for the study of quantum periods of anharmonic oscillators.  We will report more on this in \cite{GFMS}.

\acknowledgments
We would like to thank Santiago Codesido for useful discussions.
The authors thank the CERN theory division and INFN Tor Vergata for hospitality at various stages of this work.
The work of FF and JFM is partially supported by the MIUR PRIN Grant 2020KR4KN2 ``String Theory as a bridge between Gauge Theories and Quantum Gravity''. The work of AG is partially supported by the Swiss National Science Foundation Grant No.~185723 and the NCCR SwissMAP.

\begin{appendix}

  \section{SW curves for SQCD }\label{App:SCQDtoAD}

In this Appendix we review the SQCD/AD dictionary for  SW curves.
 The SW curves for a $SU(2)$ gauge theory with $0<N_f<4$ hypermultiplets transforming in the fundamental representation of the gauge group are given by
 \be
 \hat y^2 + \hat y P(x)+ q \prod_{i=1}^{N_f} (x-m_i)=0
 \ee
 with $q=\Lambda^{4-N_f}/4$ and
 \be
  P(x)=
\left\{
\begin{array}{ll}
  x^2-u \qquad& N_f=1\,,     \\
   x^2-u +q  \qquad& N_f=2 \,,    \\
    x^2-u +q ( x- \sum_i m_i ) \qquad& N_f=3   \,.  \\
\end{array}
\right.
\ee
is chosen in such a way that $u=  \frac12{\rm tr} \varphi^2$.
The  periods of the holomorphic one-form are
  \be
{\partial a(u)  \over \partial u} =   {1\over 2 \pi i}  \oint_{\alpha}  {dx\over w(x) } \qquad , \qquad
{\partial a_D(u)  \over \partial u} =     {1\over 2 \pi i}  \oint_{\beta}  {dx\over w(x) }
 \label{periods}
\ee
with
\be
 w(x)^2=d_0 \prod_{i=1}^4 (x-e_i) =\sum_{i=0}^4 d_i x^{4-i} =P(x)^2  -4 q \, \prod_{i=1}^{N_f} (x-m_i)  \label{wcurve}
 \ee
  The quantum correlator of the gauge theory can be obtained from the large $x$-expansion of the SW differential
  \be
-2\pi {\rm i}  \lambda = x {d  \log \hat y(x) \over dx} \approx \sum_{n=0}^\infty   { \left\langle   {\rm tr} \varphi^n  \right \rangle  \over x^n } \approx  2+{2u\over x^2}+\ldots
  \ee
leading to $u=  \frac12{\rm tr} \varphi^2$.  To write the  elliptic curve (\ref{wcurve}) into the Weierstrass  form, we introduce the variables $(y,z)$ related to $(w,x)$ via
\be
{1\over x-e_4} = {z\over \nu}+\delta \qquad, \qquad w={ {\rm i} \, \nu \, y \over 2(z+\nu \delta)^2}
\ee
 with
 \be
 \nu=d_0 \prod_{i=1}^3 (e_i-e_4) \qquad, \qquad \delta = \frac13 \sum{1\over e_i-e_4}
 \ee
In the new variables the SW curve takes the Weierstrass form
\be
y^2=4 z^3-g_2 z-g_3 \label{weier}
\ee
with
 \bea
g_2 &=&  \frac{4 d_2^2}{3}-4 d_1 d_3+16 d_0 d_4 \nn\\
g_3 &=& \frac{8 d_2^3}{27}-\frac{4}{3} d_1 d_3 d_2-\frac{32 d_0 d_2 d_4}{3}+4 d_3^2 d_0+4 d_1^2 d_4
 \label{g2g3}
\eea
Finally the discriminant of the Weierstrass is given by
\be
\Delta = 16 (g_2^3-27 g_3^2) \label{delta}
 \ee

 \subsection{ $\mathcal{H}_0$ theory}

 The AD $\mathcal{H}_0$ theory can be obtained  by tuning the parameters spanning the moduli space of $SU(2)$ with $N_f=1$ fundamental flavors.
 For $N_f=1$, the elliptic curve  is given by
 \be
 y^2=4 z^3- g_2 z -g_3
 \ee
 with
 \bea
 g_2 &=& \frac{64 u^2}{3} + 16 \Lambda ^3 m\nn\\
 g_3&=&\frac{512 u^3}{27} +\frac{64}{3} \Lambda ^3 m u+ 4 \Lambda ^6
 \eea
The AD point is obtained by taking
\bea
u &=& {3 \Lambda^2\over 4} +u_{AD}\, {\Lambda^{4\over 5} \over 4} - c_{AD} \, {\Lambda^{6\over 5} \over 4} \qquad , \qquad m=
-{3 \Lambda\over 4} + c_{AD} \, {\Lambda^{1\over 5} \over 4} \nn\\
 z  &=& \tilde z \Lambda^{8\over 5}  \qquad , \qquad    y  = \tilde y \Lambda^{12\over 5}
\eea
and keeping the leading order in $u_{AD},c_{AD}\to0$. This leads to
\be
\tilde y^2 =4 \tilde z^3+ 4c_{AD} \tilde z  - 4 u_{AD}
\ee
The same SW curve can be obtained from the quartic expression
\be
w^2=z^8 \left( {1\over z^7} + {c_{AD}\over z^5} + {u_{AD} \over z^4} \right)\,,
\ee
where the SW differential is given by \cite{Gaiotto:2009we} (see also \cite{Tachikawa:2013kta} for a review)
\be
\lambda=  \frac{w}{z^4} {\rm d}z
\ee

 \subsection{ $\mathcal{H}_1$ theory}

 The AD $\mathcal{H}_1$ theory can be obtained  by tuning the parameters spanning the moduli space of $SU(2)$ with $N_f=2$ flavors transforming in the fundamental representation of the gauge group. The elliptic curve  is given by
 \be
 y^2=4 z^3- g_2 z -g_3
 \ee
  with
 \bea
 g_2 &=&  \frac{4}{3} \left(\Lambda ^2 \left(\Lambda ^2+12 \mu ^2-12 m^2\right)+16 u^2-4 \Lambda ^2
   u\right) \nn\\
 g_3&=& \frac{8}{27} \left(\Lambda ^4 \left(\Lambda ^2+18 \mu ^2+36 m^2\right)-6 \Lambda ^2 u
   \left(\Lambda ^2-12 \mu ^2+12 m^2\right)+64 u^3-24 \Lambda ^2 u^2\right)\nn
 \eea
 with $m= \frac12(m_1+m_2)$, $\mu= \frac12(m_1-m_2)$.
The AD point is obtained by taking
\bea
u &=&\frac{\Lambda ^2}{2} +  \Lambda^{2\over 3}\,u_{AD} +\Lambda
    \mu - \frac{\Lambda ^{4\over 3} }{4} c  \qquad , \qquad m=
\frac{\Lambda }{2}-\frac{ \Lambda^{1\over 3}  }{4} c \nn\\
 z  &=& \tilde z \Lambda^{4\over 3}  \qquad , \qquad    y  = \tilde y \Lambda^{2}
\eea
and taking $u_{AD}$, $c$ and $\mu$ small
\be
\tilde y^2 =4 \tilde z^3- 4\left( 4 u_{AD}+{c^2\over 3} \right) \tilde z- 4 \left(  \frac{2}{27}  c^3-  \frac83 c\,  u_{AD}  +\mu ^2 \right)
\ee
  The same curve is obtained starting from the standard $\mathcal{H}_1$ quartic form
\be
w^2 = z^8 \left( {1\over z^8} + {c\over z^6} +{\mu\over z^5}+ {u_{AD} \over z^4} \right)
\ee

 \subsection{ $\mathcal{H}_2$ theory}

 The AD $\mathcal{H}_2$ theory can be obtained  by tuning the parameters spanning the moduli space of $SU(2)$ with $N_f=3$ flavors transforming in the fundamental representation of the gauge group. The elliptic curve  is given by
 \be
 y^2=4 z^3- g_2 z -g_3
 \ee

 \bea
 g_2 &=&\frac{64 u^2}{3}{+}\frac{8\Lambda}{3}   \left(2 C_3{-}3 C_2 m{+}6 m
   \left(m^2{+}u\right)\right){+}\Lambda ^2 \left(C_2{+}6 m^2{-}\frac{4 u}{3}\right){-}\frac{\Lambda ^3 m}{2}
   +\frac{\Lambda ^4}{192}\nn\\
 g_3&=&\frac{512 u^3}{27} {+}\frac{32}{9} \Lambda  u \left(2 C_3{+}6 m \left(m^2{+}u\right){-}3 C_2 m \right)-\frac{\Lambda ^5 m}{96}   +\frac{\Lambda
   ^6}{13824} \nn\\
&& +\Lambda ^2 \left(C_2^2-4 C_2
   m^2-\frac{16 C_3 m}{3}-\frac{8 C_2 u}{3}+20 m^4+\frac{20 u^2}{9}\right) \\
 &&  +\frac{ \Lambda ^3 }{18} \left(-21 C_2 m+2 C_3-30 m^3+30 m u\right) +\frac{\Lambda ^4  }{144} \left(3 C_2+54 m^2-4
   u\right) \nn
 \eea
 with
 \be
 m= \frac13 \sum_i m_i \qquad , \qquad  C_2= \sum_i (m_i-m)^2 \qquad   C_3= \sum_i (m_i-m)^3
 \ee
The AD point is obtained by taking
\bea
u &=&\frac{5 \Lambda
   ^2}{64}- \Lambda^{1\over2} \left( u_{AD} +{c^3  \over 24} \right) +{3   \Lambda^{3\over 2} \over 16} c + { \Lambda \over 16} c^2 \qquad , \qquad m=-\frac{\Lambda }{8}  -\frac{\Lambda^{1\over 2 }}{4} c \nn\\
 z  &=& \tilde z  \Lambda  \qquad , \qquad    y  = \tilde y \Lambda^{3\over 2}
\eea
and taking $u_{AD}$, $c$ and $C_2,C_3$ small
\be
\tilde y^2 =4 \tilde{z}^3+ \tilde{z}  (4c u_{AD}+{c^4\over 12}-2 C_2 )-4 u_{AD}^2+ \frac{c^6}{432}-\frac{c^2 C_2}{6}-\frac{4 C_3}{3}\ee
 The same curve is obtained starting from the standard $\mathcal{H}_2$ quartic form
\be
w^2 = z^6 \left( {1\over z^6} + {c\over z^5} +{\mu\over z^4}+ {u_{AD} \over z^3}+{M^2\over z^2} \right)
\ee
  after the identification
  \bea
 C_2 &=&  \frac{1}{24} \left(c^4+16 \mu ^2+192 M^2-8\mu c^2\right) \nn\\
 C_3 &=&  \frac{1}{288} \left(64 \mu ^3-c^6-24 c^2 \mu ^2+576 c^2 M^2-2304 \mu  M^2+12c^4\mu-24c^2\mu^2\right)
\eea

 \section{Modular functions}\label{EisensteinApp}

  In this section we collect some definitions and useful modular identities.
 The Eisenstein series are defined as
 \bea
 E_{k}(q) =1+{2\over \zeta(1-k)} \sum_{n=1}^\infty  { n^{k-1} q^{2n}\over 1-q^{2n}}
 \eea
 A basis of modular forms is given by $E_4$, $E_6$ and the quasi-modular form $E_2$. They  are related to the theta functions via
 \bea
E_4 &=&  \frac12(\theta_2^8+\theta_3^8+\theta_4^8) \nn\\
E_6^2&=&  \frac18 \left[ (\theta_2^8+\theta_3^8+\theta_4^8)^3-54 \, 2^8 \, \eta^{24}  \right] \nn\\
E_2 &=& 12 q \partial_q  \log \eta(q) \label{es}
\eea
 We introduce the functions
 \be
 K_2=\theta_3^4+\theta_4^4 \qquad , \qquad  L_2=\theta_2^4
 \ee
 Under $S$-duality they transform as
 \bea\label{SdualityKL}
 K_2(-1/\tau) &=& -\tau^2 {K_2(\tau) +3L_{2}(\tau) \over 2} \nn\\
 L_2(-1/\tau) &=& -\tau^2 {K_2(\tau) -L_{2}(\tau) \over 2}
 \eea
 whereas under $T$-duality they transform as
 \bea\label{TdualityKL}
 K_2(\tau+1) &=& K_2(\tau)  \nn\\
 L_2(\tau+1) &=& -L_{2}(\tau)
 \eea
 In terms of these variables the Eisenstein series  read
 \be
 E_4 = {K_2^2+3 L_2^2\over 4} \qquad, \qquad  E_6 = {K_2 (K_2^2-9 L_2^2)\over 8}
 \ee

 \section{ ${\bf c}$-deformation of $\mathcal{H}_1$ }\label{cDefH1}

   In this Appendix we work at $\beta=1$ and consider a deformation of $\mathcal{H}_1$ obtained by turning on the IR-relevant coupling $c$. The SW curve is now described by a cubic curve with
   \be
    g_2=u \qquad ,\qquad g_3=c \,u -4 c^3
    \ee
    and discriminant
    \be
    \Delta=16(u-3 c^2)(u-12 c^2)^2
    \ee
  We notice that we have now two monopole points $u=3 c^2$ and $u=12 c^2$.
  It is convenient in this case to introduce the modular functions $K_2$ and $L_2$ related to $E_4$ and $E_6$ via
 \be
 E_4 = {K_2^2+3 L_2^2\over 4} \qquad, \qquad  E_6 = {K_2 (K_2^2-9 L_2^2)\over 8}
 \ee
  Plugging this into (\ref{unu}) and solving for $ \omega_1$ and $u$ one finds three inequivalent solutions   \bea\label{SolutionsKL}
  u=3c^2\left(1+{ 3 L_2^2\over K_2^2} \right)\,, \qquad\qquad &&   \omega_1 = {\rm i} \sqrt{K_2\over 3c}\nn \\
 u= \frac{12 c^2 \left(K_2^2+3 L_2^2\right)}{(K_2-3 L_2)^2} \,, \qquad\qquad &&  \omega_1={\rm i} \frac{\sqrt{3
L_2-K_2}}{\sqrt{6} \sqrt{c}} \nn\\
u= \frac{12 c^2 \left(K_2^2+3 L_2^2\right)}{(K_2+3
L_2)^2} \,, \qquad\qquad &&  \omega_1= -{\rm i}\frac{\sqrt{-K_2-3 L_2}}{\sqrt{6} \sqrt{c}}\,.
  \eea
  These solutions correspond to three different duality frames related by $S,T$ transformations \eqref{SdualityKL}, \eqref{TdualityKL}. In order to make contact with the results derived in Sec.~\ref{Sec:H1b0} for the mass-deformed $\mathcal{H}_1$ theory, we choose the duality frame given by the second line in \eqref{SolutionsKL}.
 To make formulae simpler, it is convenient to define
 \be
 \hat{K}_2\equiv-\frac{K_2-3L_2}{2}\qquad,\qquad \hat{L}_2\equiv-\frac{K_2+L_2}{2}\,,
 \ee
 which correspond to the ST transformations of $K_2$ and $L_2$ respectively.
 This leads to
  \be
  {\cal F}_1={1\over 12} \log \left( c^9  { \hat{L}_2^2 (\hat{L}_2^2-\hat{K}_2^2)^2 \over \hat{K}_2^9 } \right)   \qquad , \qquad \xi={2\over {\rm i}}{\hat{K}_2^{5\over 2} \over \sqrt{27 c^3} \hat{L}_2^2(\hat{K}_2^2-\hat{L}_2^2)}
  \ee
  Likewise we define $u_D$, $\omega_{1D}=da_D/du$, ${\cal F}^D_g $ by the same formulae \eqref{SolutionsKL} replacing $K_2\to\hat{K}_2$ and $L_2\to\hat{L}_2$.

The holomorphic ambiguity for the theory has the following form (see \cite{Codesido:2017dns})
\be h_g=\sum _{i=0}^{3 g-4} \hat{L}_2^{2 i} \hat{K}_2^{3 (g-1)-2 i} h_{g,i}\ee
with coefficients $h_{g,i}$ determined by requiring that both ${\cal F}_g $ and ${\cal F}^D_g $ satisfy the gap conditions when $a\to 0$ and $a_D\to 0$ respectively, i.e. $q\to 0$ or $q_D \to 0$.
Solving recursively, the holomorphic anomaly equation (\ref{HAE})  one finds the first few terms
\bea
&&{\cal F}_2 ={ \xi^2\over  24^3}\left(\frac{5 E_2^3}{3}+\frac{3 E_2^2 \left(\hat{K}_2^2+\hat{L}_2^2\right)}{2 \hat{K}_2}+E_2 \left(-\frac{81 \hat{L}_2^4}{4 \hat{K}_2^2}-\frac{55 \hat{K}_2^2}{4}+15 \hat{L}_2^2\right)+h_2(q)\right)\nn\\
&&{\cal F}_3 ={ \xi^4\over 24^5}\left(\frac{5 E_2^6}{6}+E_2^5 \left(5 \hat{K}_2-\frac{20 \hat{L}_2^2}{3 \hat{K}_2}\right)+E_2^4 \left(\frac{59 \hat{L}_2^4}{8 \hat{K}_2^2}+\frac{83 \hat{K}_2^2}{8}-\frac{263 \hat{L}_2^2}{12}\right)\right.\nn\\
&&+E_2^3 \left(\frac{447 \hat{L}_2^6}{8 \hat{K}_2^3}-\frac{403 \hat{K}_2^3}{18}-\frac{77 \hat{L}_2^4}{2 \hat{K}_2}+\frac{465 \hat{K}_2 \hat{L}_2^2}{8}\right)+h_3(q)\nn\\
&&+\frac{E_2 \left(199822 \hat{K}_2^8-779495 \hat{K}_2^6 \hat{L}_2^2+1133751 \hat{K}_2^4 \hat{L}_2^4-801225 \hat{K}_2^2 \hat{L}_2^6-376245 \hat{L}_2^8\right)}{480 \hat{K}_2^3}\nn\\
&&\left.-\frac{E_2^2 \left(77573 \hat{K}_2^8-254926 \hat{K}_2^6 \hat{L}_2^2+292815 \hat{K}_2^4 \hat{L}_2^4-129600 \hat{K}_2^2 \hat{L}_2^6+65610 \hat{L}_2^8\right)}{480 \hat{K}_2^4}\right)
\eea
where
\bea h_2&=&\frac{1619 \hat{K}_2^3}{120}-\frac{279 \hat{L}_2^4}{8 \hat{K}_2}-\frac{111 \hat{K}_2 \hat{L}_2^2}{4}\\
h_3&=&-\frac{11660261 \hat{K}_2^6}{40320}-\frac{3753 \hat{L}_2^{10}}{8 \hat{K}_2^4}+\frac{20885 \hat{K}_2^4 \hat{L}_2^2}{16}-\frac{303615 \hat{L}_2^8}{128 \hat{K}_2^2}-\frac{733469 \hat{K}_2^2 \hat{L}_2^4}{320}\nonumber\\&&+\frac{31887 \hat{L}_2^6}{16} \eea
  These are such that
  \bea
  {\cal F}_g^D &=&  (-1)^{g-1}2^{2g-1}\frac{ B_{2 g}}{2 g(1-g)}{1\over a_D^{2g-1}}+\mathcal{O}(a_D^0)\nn\\
   {\cal F}_g &=& (-1)^{g-1} 2^{2g-2}\frac{ B_{2 g}}{2 g(1-g)}{1\over a^{2g-1}}+ \mathcal{O}(a^0)\,.
   \eea
   %{\color{red}CONTROLLARE/CAMBIARE}
   %where the extra factor of $2$ in the second line comes from the fact that  the discriminant has a zero of order two at $u=12 c^2$, so two particles are becoming massless at this point.

     \subsection*{The conformal limit}

     This is a theory of type $\text{\bf B}$, hence $\tau^*={\rm i}$ at the conformal point. In particular at this point both $L_2$ and $K_2$ are finite with
     \be K_2\Big|_{\tau={\rm i}}=3 L_2\Big|_{\tau={\rm i}}, \qquad L_2\Big|_{\tau={\rm i}}=3^{1/2} \sqrt{E_4}\Big|_{\tau={\rm i}}\ee
  and \be a\approx \frac{32 i \sqrt{2} c^{3/2} K_2^2}{27 \left(L_2-\frac{K_2}{3}\right)^{3/2}}+\frac{4 i \sqrt{2} c^{3/2} (6 E_2+4 K_2)}{9 \sqrt{L_2-\frac{K_2}{3}}}+O\left(\sqrt{L_2-\frac{K_2}{3}}\right) \ee
  as well as      \be\ba  {\cal F}_2&\approx -\frac{E_2}{96 a^2}~,\qquad
  {\cal F}_3\approx \left(-\frac{E_2^2}{1152}-\frac{139E_4}{34992}\right){1\over a^4}~, \quad  {\cal F}_4\approx ~-\frac{E_2 \left(7533 E_2^2+106752 E_4\right)}{40310784 a^6},\\
 \qquad
  {\cal F}_5&\approx\frac{-597051 E_2^4-17173728 E_2^2 E_4-44454429 E_4^2}{8707129344 a^8} ~, \dots
      \ea\ee
which agrees with the results of Sec.~\ref{Sec:H1b0}. This provides an explicitly test that the conformal limit is independent of the deformation we perform.

 \section{ SQCD$_{\bf N_f=3}$ at the conformal point }\label{SQCDNf3}

      The SW curve of SQCD with $N_f=3$ flavors of equal mass $m=-{\Lambda\over 8} $ is described by a curve
   in the Weirstrass form with
  \bea
 g_2&=& \frac{1}{192} \left(64 u-5 \Lambda ^2\right)^2 \qquad  , \qquad  g_3= \frac{\left(64 u-5 \Lambda ^2\right)^2 \left(17 \Lambda
   ^2+128 u\right)}{27648}\nn\\
   \Delta &=& -\frac{\Lambda ^2 \left(64 u-5 \Lambda ^2\right)^4
   \left(7 \Lambda ^2+256 u\right)}{65536}
  \eea
   For this choice, formula (\ref{unu}) can be explicitly solved and one finds
  \bea
 u&=&  -\frac{\Lambda ^2 \left(17\, E_4^{3/2}+10
   E_6\right)}{128
   \left(E_4^{3/2}-E_6\right)} \qquad, \qquad
%    \Delta=\frac{3^{15}  \Lambda ^{12} E_4^6
%   \left(E_4^{3/2}+E_6\right)}{4^{10}
%    \left(E_4^{3/2}-E_6\right)^5} \nn\\
\xi =   \frac{16
   \sqrt{  \frac{2}{3} \left( E_4^{3/2}-E_6 \right) }}{3 \Lambda (
   E_4^{3/2}+  E_6) }
  \eea
    and
    \be
\mathcal{F}_1 = \frac{1}{12} \log \left(\frac{\Lambda ^{18}
   E_4^9
   \left( E_4^{3/2}+E_6\right)}{\left(
  E_4^{3/2}-E_6\right)^8}\right)
     \ee
Plugging this into the anomaly equation, one finds for the first few gravitational corrections at $\beta=1$
{\small
  \bea
 && {\cal F}_2  = {\xi^2\over 24\, 12^2 }  \left[\frac{5
  E_2^3}{3} {+}\frac{3 E_2^2 \left( 3 E_4^{3/2}{+}4
   E_6\right)}{
 E_4}{-}\frac{E_2 \left({-}36
 E_4^{3/2} E_6{+}7 E_4^3{+}12
  E_6^2\right)}{ E_4^2}{+} h_2(q) \right] \nn\\
 &&  {\cal F}_3 = {\xi^4\over 24\, 12^4} \left[
   \frac{5 E_2^6}{6}{+}\left(\frac{5 E_6}{E_4}{-}5 E_4^{1\over 2}\right) E_2^5{-}\left(\frac{8 E_6^2}{E_4^2}{+}\frac{49
   E_6}{E_4^{1\over 2}}{-} \frac{E_4}{2}\right) E_2^4\right. \nn\\
   && \left.
   {+}\left(\frac{68 E_6^3}{3 E_4^3}{+}\frac{48 E_6^2}{E_4^{3\over 2}}{-}\frac{520
   E_6}{9}{+}96 E_4^{3\over 2}\right) E_2^3{-}\left(\frac{48 E_6^4}{E_4^4}{+}\frac{144 E_6^3}{E_4^{5\over 2}}{+}\frac{482
   E_6^2}{E_4}{-}\frac{7108}{5} E_4^{1\over 2}  E_6{+}\frac{4669 e_4^2}{6}\right) E_2^2 \right.\nn\\
 && \left.
   {+}\left(\frac{12 E_6^3}{E_4^2}-\frac{9676
   E_6^2}{5 E_4^{1\over 2}} +\frac{101773 E_4 E_6}{15}-\frac{22947 E_4^{5\over 2}}{5}\right) E_2+h_3(q)
    \right] \nn\\
    &&  {\cal F}_4 = {\xi^6\over 24\, 12^6}\left[\frac{1105 E_2^9}{1296}+E_2^8 \left(\frac{5 E_6}{E_4}-\frac{625 \sqrt{E_4}}{48}\right)+\frac{E_2^7 \left(-3543 E_4^{3/2} E_6+3130 E_4^3-270 E_6^2\right)}{36 E_4^2}\right.\nn\\
    && +\frac{1}{108} E_2^6 \left(\frac{9912 E_6^2}{E_4^{3/2}}-31530 E_4^{3/2}+\frac{2448 E_6^3}{E_4^3}+73105 E_6\right)\nn\\
    &&+ \frac{E_2^5 \left(128520 E_4^{3/2} E_6^3+232890 E_4^{9/2} E_6+172505 E_4^6+235218 E_4^3 E_6^2+27480 E_6^4\right)}{360 E_4^4}\nn\\
    &&+\frac{E_2^4 \left(442080 E_4^{3/2} E_6^4+2633742 E_4^{9/2} E_6^2+5296335 E_4^{15/2}-7947226 E_4^6 E_6+939180 E_4^3 E_6^3+94080 E_6^5\right)}{360 E_4^5}+\nn\\
&&    \frac{E_2^2 \left(2372448 E_4^{3/2} E_6^4-696242070 E_4^{9/2} E_6^2-897505518 E_4^{15/2}+1568619971 E_4^6 E_6+52768708 E_4^3 E_6^3+733600 E_6^5\right)}{1260 E_4^4}\nn\\
    &&-\frac{E_2^3}{540 E_4^6}{\left(1555200 E_4^{3/2} E_6^5+2157384 E_4^{9/2} E_6^3-54053697 E_4^{15/2} E_6+27173594 E_4^9+20978368 E_4^6 E_6^2\right.}\nn\\
    && \left. \qquad\qquad+3306360 E_4^3 E_6^4+311040 E_6^6\right)\nn\\
  &&  -\frac{E_2}{25200 E_4^5}{\left(512517120 E_4^{3/2} E_6^5-3951672720 E_4^{9/2} E_6^3-95049030780 E_4^{15/2} E_6+53185189825 E_4^9\right.}\nn\\
  &&\qquad \qquad\quad \left.+49201422412 E_4^6 E_6^2+1174650320 E_4^3 E_6^4+92288000 E_6^6\right)\nn \\
  &&\left.+h_4(q)
   \right]
    \eea
  }
 The ambiguous part is given by
 \bea
h_2 & =&   \frac{1106 E_6}{15}- \frac{171 E_4^{3\over 2}}{5} \nn\\
h_3 & =& - \frac{1648 E_6^4}{9 E_4^3}- \frac{24144 E_6^3}{35 E_4^{3/ 2}}- \frac{1234978 E_6^2}{315}+ \frac{292119}{35} E_4^{3 \over 2}
   E_6- \frac{850279 E_4^3}{126} \nn\\
 h_4 & =&   \frac{795392 E_6^7}{27 E_4^6}+ \frac{2311168 E_6^6}{15 E_4^{9/2}}+ \frac{13813852 E_6^5}{45 E_4^3}+ \frac{79514072
   E_6^4}{315 E_4^{3/2}}+ \frac{9197665261 E_6^3}{28350} \nn\\
   && - \frac{3057458963 E_4^{3/2} E_6^2}{1260}+ \frac{15201332353 E_4^3
   E_6}{3780}- \frac{3669761651 E_4^{9/2}}{1680}
 \eea

   \subsection*{The conformal limit}
   This is a theory of type $\text{\bf A}$, hence $\tau^*=e ^{\pi {\rm i} \over 3}$.
  By perturbing around this point we get
  \be a \approx\frac{9 \sqrt{3}E_4}{8 \sqrt{E_6}}  \ee
  and
  \be \ba
  \mathcal{F}_2&\approx-\frac{E_2}{128 a^2}\qquad\qquad\qquad\qquad\qquad
    \mathcal{F}_3\approx-\frac{E_2^2}{2048 a^4}\\
      \mathcal{F}_4&\approx\frac{3107 E_6}{663552 a^6}-\frac{3 E_2^3}{32768 a^6}\qquad\qquad
        \mathcal{F}_5 \approx\frac{34177 E_2 E_6}{5308416 a^8}-\frac{97 E_2^4}{3145728 a^8}\\
        \vdots
  \ea\ee
  These can be resummed using the hypergeometric functions  \eqref{zu}  and in agreement with \eqref{csH2}.   This provides an explicitly test that the conformal limit is independent of the deformation we perform.

  \section{Holomorphic ambiguities for $\mathcal{H}_0,\mathcal{H}_1,\mathcal{H}_2$ at $\beta=1$}
 \label{AppE}
  The holomorphic anomaly equation determines $\widehat{\cal Z}(a,\beta)$ up to the $E_2$-independent part of \eqref{zu}, i.e.
  \be
 \widehat{\cal Z}(a,\beta)|_{E_2\to0} =  \sum_{n=0}^{\infty}c_n\left(-\frac{\epsilon_1\epsilon_2}{6a^2}\right)^{n\delta-\tfrac{\gamma}2}E_{2\delta}^n\,.
\ee
In this Appendix we list the first few $c_n$ coefficients at $\beta=1$ for the three AD theories we analyzed in this paper.

  \subsection*{$\bf \mathcal{H}_0$}

  \bea \label{cnH0} c_2&=& -\frac{3411230845030961039}{2^{17} 3^2 5^{14}}\,,\nn\\
 c_3&=& -\frac{11228416395151247860243314067849}{2^{25} 3^4 5^{21}}\,,\nn\\
  c_4&= &-\frac{336921369293660561201677735133941404089137439}{2^{35} 3^5 5^{28}}\,,\nn\\
  c_5&= &-\frac{54446679876958884558177953879909686803701101902116733352249}{2^{43} 3^6 5^{36}}\,.
\eea
These coefficients determine the behavior of ${\mathfrak{f}_g}$ in  \eqref{Fad} up to $g=18$.

  \subsection*{$\bf \mathcal{H}_1$}

   \bea\label{cnH1}
   c_2=& -\frac{399471589}{60466176}\,,\nn\\
   c_3=&-\frac{231844286893415}{176319369216}
   \eea
 These coefficients determine the behavior of ${\mathfrak{f}_g}$ in  \eqref{Fad} up to $g=7$.

    \subsection*{$\bf \mathcal{H}_2$}

        \bea\label{cnH2}
    c_2&=&-\frac{23495274215}{2^{21} 3^2}\,, \nn \\
    c_3&=&-\frac{4120670292728086475}{2^{31} 3^4}\,,  \nn\\
    c_4&=&-\frac{6480114817503034769242602575}{2^{43} 3^5}\,.
    \eea
  These coefficients determine the behavior of ${\mathfrak{f}_g}$ in  \eqref{Fad} up to $g=15$.

 \end{appendix}

 \providecommand{\href}[2]{#2}\begingroup\raggedright\endgroup
 
\end{document}